\documentclass[
reprint,
superscriptaddress,
nofootinbib,
amsmath,amssymb,
aps,
prstab,
floatfix,
]{revtex4-2}

\usepackage{graphicx}
\usepackage{dcolumn}
\usepackage{bm}

\usepackage[T1]{fontenc}
\usepackage{newtxmath,newtxtext}
\usepackage{mathtools}
\usepackage{physics}
\usepackage{enumitem}
\usepackage[caption=false]{subfig}
\usepackage{derivative}
\usepackage{natbib}
\usepackage{cancel}
\usepackage{xcolor}
\usepackage{xr-hyper}
\usepackage{hyperref}

\usepackage{ragged2e}
\DeclareCaptionJustification{justified}{\justifying}

\makeatletter
\def\bibsection{%
	\par
	\begingroup
	\baselineskip26\p@
	\bib@device{\hsize}{72\p@}%
	\endgroup
	\nobreak\@nobreaktrue
	\addvspace{19\p@}%
}%
\makeatother

\begin{document}
	\preprint{APS/123-QED}
	
	\title{Impact of bunch intensity asymmetry in colliders featuring strong beamstrahlung}

	\author{Khoi Le Nguyen Nguyen}
	\email{kl518@cam.ac.uk}
	\affiliation{European Organisation for Nuclear Research (CERN), CH 1211 Geneva 23, Switzerland}
	\affiliation{\'Ecole Polytechnique F\'ed\'erale de Lausanne (EPFL), Route Cantonale, 1015 Lausanne, Switzerland}
	\affiliation{Cavendish Laboratory, University of Cambridge, JJ Thomson Avenue, Cambridge CB3 0HE, UK}
	
	\author{Xavier Buffat}
	\email{xavier.buffat@cern.ch}
	\affiliation{European Organisation for Nuclear Research (CERN), CH 1211 Geneva 23, Switzerland}
	
	\author{Peter Kicsiny}
	\email{peter.kicsiny@cern.ch}
	\affiliation{European Organisation for Nuclear Research (CERN), CH 1211 Geneva 23, Switzerland}
	\affiliation{\'Ecole Polytechnique F\'ed\'erale de Lausanne (EPFL), Route Cantonale, 1015 Lausanne, Switzerland}
	
	\author{Tatiana Pieloni}
	\email{tatiana.pieloni@epfl.ch}
	\affiliation{\'Ecole Polytechnique F\'ed\'erale de Lausanne (EPFL), Route Cantonale, 1015 Lausanne, Switzerland}
	
	\date{\today}
	
	\begin{abstract}
		An analytical investigation of beamstrahlung-induced blow-up in Gaussian beams with arbitrary dimensions is presented, using various approximations for the strength of the hourglass effect and crab waist scheme. The results, applied to the FCC-ee resonances, are compared with simulations and previous calculations, and relative luminosity values are also calculated. The stability of resultant conformations are analysed to rule out the existence of a flip-flop phenomenon in the longitudinal plane analogous to the well-known transverse counterpart. Implications for the top-up injection procedures are discussed, and a phenomenological model is proposed to study the transverse-longitudinal coupling in the blowup dynamics.
	\end{abstract}
	\maketitle

	\section{Introduction}	
	In high-energy electron-positron circular colliders, particles emit synchrotron radiation when their trajectories are bent by external magnetic fields~\cite{Wiedemann} and by the electromagnetic field generated by the opposing beam at the Interaction Point (IP)~\cite{Chen}. In the next generation of machines such as the proposed Future Circular Collider (FCC), the strength of the latter, so-called beamstrahlung, leads to a variety of new phenomena deteriorating the beams quality and thus the luminosity~\cite{Shatilov}. One consequence of beamstrahlung is the increase of the bunch length many times above the equilibrium given by the synchrotron radiation in the magnetic lattice only, thereby increasing the energy spread as well as lowering the luminosity. This effect has been studied analytically for the case of identical machine and beam parameters for the two rings~\cite{Garcia}. While configurations featuring asymmetric machine or beam parameters can be addressed with numerical simulations featuring beamstrahlung ~\cite{Shatilov}, analytical methods to rapidly approximate equilibrium lengths in the general case are not available.
	
	In the presence of an asymmetry in the intensity of two colliding bunches, an instability mechanism analogous to the flip-flop effect \cite{Tennyson}, termed \textit{3D flip-flop}, was observed in simulations~\cite{Shatilov}: As the electromagnetic fields generated by the lowest intensity bunch are reduced, the beamstrahlung of the highest intensity bunch is reduced. Consequently, the bunch length of the high intensity bunch is reduced, thus increasing the strength of the beam-beam force exerted on the low intensity bunch. This results in an increase of the beamstrahlung of the low intensity bunch, thus increasing its length and further weakening the beam-beam force on the strong bunch. Beyond a certain threshold the beam-beam force experienced by the low intensity bunch becomes too strong, such that non-linear resonances drive a transverse emittance growth that again weakens the low intensity bunch, thus enhancing the phenomenon. We obtain a runaway situation in which the low intensity bunch blows up in all directions while the high intensity bunch shrinks to the equilibrium bunch properties defined by the magnetic lattice, as the beamstrahlung due to the blown up low intensity bunch becomes negligible. 
	
	This paper presents an analytical treatment for the effect of beamstrahlung on the equilibrium bunch lengths of two colliding Gaussian electron-positron bunches with arbitrary dimensions and intensities. The dispersive effects at the IP are ignored. Beam losses due to the limited energy acceptance of the ring are also ignored. Corrections due to the hourglass effect and the crab-waist collision scheme will be considered up to first order. At first only variations of the longitudinal bunch dimension are considered to obtain luminosity values and equilibrium bunch behaviours at different mean intensities and intensity asymmetry. The results obtained are compared to past studies for symmetric configurations. The stability of the solutions is then discussed. The effect of beamstrahlung on the vertical size of the low intensity bunch is then accounted for based on a phenomenological approach, leading to a model of the transverse-longitudinal flip-flop effect. 
 
	The presented model has a predictive power limited to small, but realistic parameter asymmetries. Therefore the main focus of the paper is an attempt to use analytical derivations to understand the flip-flop effect, which has so far been demonstrated only in complicated, numerically heavy tracking simulations. It can be useful to build confidence in simulation results in scenarios, such as the FCC, where experimental data cannot be obtained before the machine is built.
	
	\section{\label{eqmlength}Equilibrium bunch lengths}
	In the following, we describe the motion of a particle in the so-called weak bunch through the force of the so-called strong bunch. At this stage of the derivation the approach is thus `weak-strong', \emph{i.e.} the perturbation of the beam-beam force during the interaction is neglected. At a later stage, the weak-strong equations derived will be combined into a system of equations where the low and high intensity bunch take the role of the weak and strong beam alternatively. The system of equations thus obtained will be effectively `strong-strong', i.e the perturbations of the two beams by the beam-beam force are taken into account self-consistently. 
	
	Consider two ultra-relativistic Gaussian bunches of electrons and positrons moving towards each other with relativistic Lorentz factors $\gamma_{w/s}$. (Subscripts $w$ and $s$ throughout this paper denote quantities pertinent to the weak and strong bunch respectively.) In the lab frame, the position of a charge in either bunch can be parameterised by three fixed Cartesian coordinates $x$, $y$ and $s$ with the origin at the centre of the weak bunch, and a comoving coordinate $z = s-ct$, such that $s$ points along the axis of the weak bunch towards the strong bunch, $x$ and $y$ are the transversal directions in and perpendicular to the plane of collision respectively, and $z$ measures the deviation of a particle of the weak bunch from its centre. A charged particle in one beam moving through the electromagnetic field of the other beam will accelerate and emit radiation, known as beamstrahlung; this exists in addition to the synchrotron radiation emitted due to the bending of the trajectories in the magnetic lattice. There exists a Lorentz transformation~\cite{Hirata} to an inertial frame of reference in which the beams collide head-on; the energies of the beams as well as of the emitted photons can be shown to only be slightly different from that measured in the lab frame for a small crossing angle~\cite{Garcia}, and this small deviation will be ignored. In this transformed frame, the electromagnetic field is now in the collision plane and perpendicular to the direction of motion.
	
	Following derivations from Refs.~\cite{Garcia, Ziemann}, the time-dependent curvature, to the power of some positive integer $n$ and averaged over the entire weak bunch, of the path of a charge influenced by the electromagnetic field of a Gaussian strong bunch in the transformed frame is given by the integral,
	\begin{align}
		\label{eqn:I-nw-general}
		&\mathcal{I}_{n,w}\equiv\int _{-\infty}^{+\infty}ds\Bigg\langle\dfrac{1}{\rho_w^n}(s,t)\Bigg\rangle\nonumber\\
		&=\dfrac{1}{(2\pi)^{3/2}}\iiiint\limits_{x,y,z,s\in\mathbb{R}^4}\dfrac{dxdydzds}{\sigma_{x,w}^*\sigma_{y,w}^*G(x,s)_{w}\sigma_{z,w}}\rho(x,y,z,s;t)_w^{-n}\nonumber\\
		&\hspace{10pt}\times\exp\left[-\dfrac{\left(x+{z\theta_c}/{2}\right)^2}{2\sigma_{x,w}^{*2}}-\dfrac{y^2}{2\sigma_{y,w}^{*2}G(x,s)_{w}^2}-\dfrac{z^2}{2\sigma_{z,w}^2}\right]\nonumber\\
		&:=\frac{1}{(2\pi)^{3/2}}\iiiint\limits_{x,y,z,s\in\mathbb{R}^4}dxdydzds\frac{\mathcal{G}}{\rho_w^n},
	\end{align}
	where the local curvature is
	\begin{align}
		\dfrac{1}{\rho(x,y,z,s;t)_w}&=\left|\mathcal{F}\left[x-\left(s-\dfrac{z}{2}\right)\theta_c,y,s\right]_s\right|\nonumber\\
		&\times\sqrt{\dfrac{2}{\pi}}\dfrac{2r_eN_s}{\gamma_w\sigma_{z,s}}\exp\left[-\dfrac{(2s-z)^2}{2\sigma_{z,s}^2}\right],
	\end{align}
	and $\mathcal{G}$ denotes the Gaussian density of the weak bunch. The bunch parameters in Eq.~\ref{eqn:I-nw-general} are taken in the Lorentz transformed, head-on reference frame. In particular, $x$ and $\sigma_{x,w}^*$ are the horizontal coordinates and r.m.s. size of a bunch tilted by the half crossing angle $\theta_c/2$. The classical electron radius is $r_e$, $N_s$ is the number of particles in the strong bunch. The factor $G(s,x)_{w/s}=\sqrt{1+[({s\mp x/\theta_c})/{\beta_y^*}]^2}$
	takes into account the dominant vertical hourglass effect (via the $s/\beta_y^*$ term) and crab waist collision scheme (via the $\pm x/(\theta_c\beta_y^*)$ term) experienced by the two beams, with optical $\beta$ function in the y-direction $\beta_y^{*}$ at the IP, and the + ($-$) sign for the strong (weak) beam respectively. The exact form of the function $\mathcal{F}(x,y,z)_s$, can be found in Refs.~\cite{Wiedemann, Garcia}, and is not reproduced here, as the focus of this paper is on machines featuring a much smaller vertical beam size at the IP with respect to the horizontal size. This feature allows for approximations that enable a fully analytical approach and will be discussed in the next section.	In Eq.~\eqref{eqn:I-nw-general}, the integral is taken over the entire interacting length, and the Gaussian kernel accounts for the particle distribution of the weak bunch. Starred quantities denote values at the IP.
	
	This integral plays an important role in controlling a variety of quantities associated with beamstrahlung ~\cite{Wiedemann, Garcia, Sands}. Here, for convenience and completeness, we reiterate these quantities. With $n=1$, Eq.~\eqref{eqn:I-nw-general} leads to the average number of photons emitted per collision
	\begin{equation}
		\label{eqn:Nph}
		\mathcal{N}_{\text{ph},w}\approx\dfrac{5}{2\sqrt{3}}\alpha\gamma_w\mathcal{I}_{1,w},
	\end{equation}
	where $\alpha$ is the fine structure constant. With $n=2$, Eq.~\eqref{eqn:I-nw-general} gives the average relative energy loss per collision due to beamstrahlung,
	\begin{equation}
		\label{eqn:deltaBS}
		\delta_{\text{BS},w}\approx\dfrac{2}{3}r_e\gamma_{w}^3\mathcal{I}_{2,w}.
	\end{equation}
	With $n=3$, Eq.~\eqref{eqn:I-nw-general} relates to the energy spread of the photons emitted and the resulting quantum excitation,
	\begin{equation}
		\label{eqn:Nphe2}
		\{\mathcal{N}_{\text{ph},w}\langle u^2\rangle\}_{z,w,\text{BS}}\approx\dfrac{55}{24\sqrt{3}}\dfrac{r_e^2\gamma_{w}^5}{\alpha}\mathcal{I}_{3,w}.
	\end{equation}
	In writing down these expressions, we have implicitly assumed that both beams have the same $\beta_y^*$ parameter. 
	\subsection{\label{subsec:evol-length}Evolution of the bunch lengths towards equilibrium}
	The rate of change of the bunch length with time is affected by the emission of synchrotron radiation in the magnetic lattice and beamstrahlung radiation at the IP. Their contributions to the damping time and the quantum excitation are labelled $SR$ and $BS$ respectively, we have:
	\begin{eqnarray}
		\odv{\sigma_{z,w}^2}{t} &=& \dfrac{1}{2}\{\dot{\mathcal{N}}_\text{ph}\langle u^2\rangle\}_{z,w,\text{SR}} + \dfrac{1}{2}\{\dot{\mathcal{N}}_\text{ph}\langle u^2\rangle\}_{z,w,\text{BS}}\nonumber\\
		&-& \left(\dfrac{2}{\tau_{z,w,\text{SR}}}+\dfrac{2}{\tau_{z,w,\text{BS}}}\right)\sigma_{z,w}^2.
	\end{eqnarray}
	We define the equilibrium bunch length due to synchrotron radiation in the arcs alone, \emph{i.e.}
	\begin{equation}
		\dfrac{1}{2}\{\dot{\mathcal{N}}_\text{ph}\langle u^2\rangle\}_{z,w,\text{SR}}=\dfrac{2}{\tau_{z,w,\text{SR}}}\sigma_{z,w,\text{SR}}^2.
	\end{equation}
	Following Ref.~\cite{Garcia}, we obtain the average energy loss due to beamstrahlung:
	\begin{align}
		&\dfrac{1}{2}\{\dot{\mathcal{N}}_\text{ph}\langle u^2\rangle\}_{z,w,\text{BS}}\approx\dfrac{2}{\tau_{z,w,\text{SR}}}\times\dfrac{n_\text{IP}\tau_{z,w,\text{SR}}}{4T_\text{rev}}\{\mathcal{N}_\text{ph}\langle u^2\rangle\}_{z,w,\text{BS}}\nonumber\\
		&\approx\dfrac{2}{\tau_{z,w,\text{SR}}}\times\dfrac{n_\text{IP}\tau_{z,w,\text{SR}}}{4T_\text{rev}}\left(\dfrac{\alpha_pC}{2\pi Q_\text{s}}\right)^2\dfrac{55}{24\sqrt{3}}\dfrac{r_e^2\gamma_{w}^5}{\alpha}\mathcal{I}_{3,w},\nonumber\\
		&\equiv\dfrac{2}{\tau_{z,w,\text{SR}}}\mathcal{A}_{w}\mathcal{I}_{3,w},
	\end{align}
	where $n_\text{IP}$ is the number of interaction points within one revolution of period $T_\text{rev}$ in a ring of circumference $C$, with momentum compaction factor $\alpha_p$ and synchrotron tune $Q_\text{s}$. For convenience, we have introduced
	\begin{equation}
		\mathcal{A}_{w}\equiv\frac{n_\text{IP}\tau_{z,w,\text{SR}}}{4T_\text{rev}}\left(\frac{\alpha_p C}{2\pi Q_\text{s}}\right)^2\frac{55}{24\sqrt{3}}\frac{r_e^2\gamma_{w}^5}{\alpha}.
	\end{equation}\\
	The beamstrahlung damping time, in units of turns, can be deduced from Eq.~\eqref{eqn:deltaBS} to be
	\begin{equation}
		\dfrac{1}{\tau_{z,w,\text{BS}}}=n_\text{IP}\delta_{w,\text{BS}}\approx\dfrac{2}{3}n_\text{IP}r_e\gamma_{w}^3\mathcal{I}_{2,w}.
	\end{equation}
	The rate of change of the bunch length can be re-written as
	\begin{align}
		\odv{\sigma_{z,w}^2}{t}&=\dfrac{2}{\tau_{z,\text{SR}}}(\sigma_{z,w,\text{SR}}^2+\mathcal{A}_{w}\mathcal{I}_{3,w})\nonumber\\
		&\hspace{10pt}-\left(\dfrac{2}{\tau_{z,\text{SR}}}+\mathcal{B}_{w}\mathcal{I}_{2,w}\right)\sigma_{z,w}^2,
		\label{eqn:evol_lengths}	
	\end{align}
	where $\mathcal{B}_{w}\equiv n_\text{IP}\frac{4}{3}r_e\gamma_{w}^3$. The equilibrium is attained when $\odv{\sigma_{z,l/h}^2}{t}=0$ leading to a set of two implicit, non-linear equations that can be solved for the equilibrium bunch length of the low and high intensity bunches, with subscripts $l$ and $h$ respectively.
	\subsection{Analytical approximations for $\mathcal{I}_{n,w}$}
\begin{figure}[h]
    		\includegraphics[width=\linewidth]{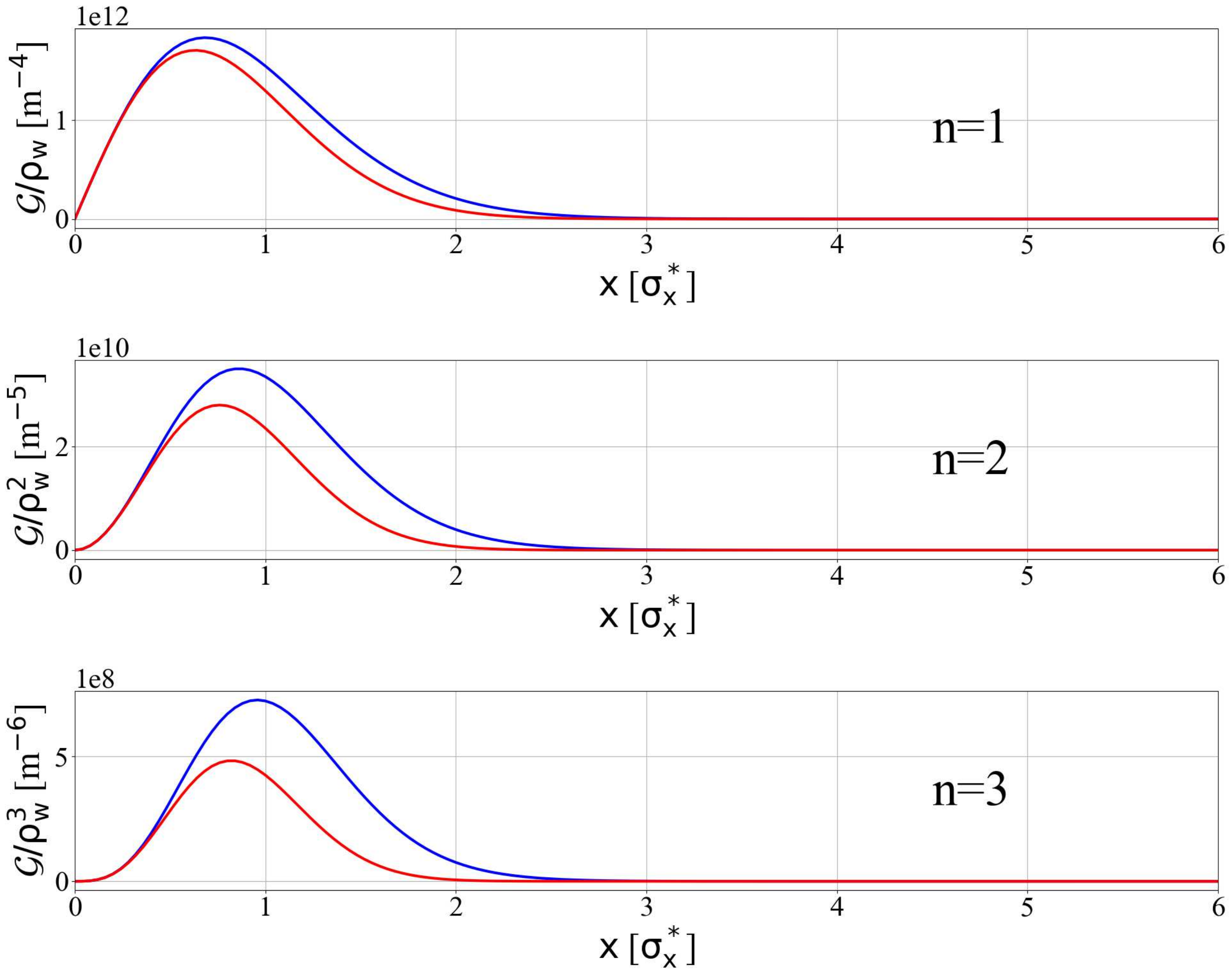}
    		\caption{Plots of the integrand of Eq.~\ref{eqn:I-nw-general}, $\mathcal{G}/\rho_{w}^n$, as a function of $x$, using the t\=t$_2$ (182.5~GeV) parameter set. Our analytical predictions (red) are compared to values obtained by simulation (blue) of a single beam-beam pass in \texttt{Xsuite}, using 101 longitudinal slices, and calculated as the ratio of the transverse kick over the longitudinal distance between two consecutive slices.}
      \label{fig:comparisons}
    	\end{figure}

	In the flat-beam approximation, $\sigma_{x,s}^*\gg\sigma_{y,s}^*$, the function $\mathcal{F}(x,y,z)_s$ can be approximated as~\cite{Garcia}
	\begin{equation}
		|\mathcal{F}(x,y,s)_s|\approx\left|\dfrac{y}{\sigma_{y,s}^*G(s,x)_s}+\dfrac{ix}{\sigma_{x,s}^*}\right|\dfrac{\exp\left[-{x^2}/({2\sigma_{x,s}^{*2}})\right]}{\sigma_{x,s}^*}.
	\end{equation}
	Considering $|x|<\sigma^*_{x,s}$ and $|y/G|<\sigma^*_{y,s}$, one may further approximate the local curvature~\cite{Garcia},
	\begin{align}
		&\frac{1}{\rho(x,y,z,s;t)_w}=\sqrt{\frac{2}{\pi}}\frac{2r_eN_s}{\gamma_w\sigma_{z,s}}\nonumber \\&\times\frac{\exp[-(x-(s-z/2)\theta_c)^2/(2\sigma_{x,s}^{*2})]}{\sigma_{x,s}^*}\exp\left[-\frac{(2s-z)^2}{2\sigma_{z,s}^2}\right]\nonumber\\
		&\times\left[\left(\frac{y}{\sigma_{y,s}^* G(s,x-(s-z/2)\theta_c)_s}\right)^2+\left(\frac{x-(s-z/2)\theta_c}{\sigma_{x,s}^*}\right)^2\right]^{1/2}. \label{eqn:local-curvature}
	\end{align}
    At this point, it is worth assessing the overall impact of the various approximations on the integrand in Eq.~\eqref{eqn:I-nw-general}. The local curvature is a function of four space variables $x$, $y$, $z$ and $s$, as well as of time $t$ (which is not an integration variable), and so a meaningful comparison between the approximation of Eq.~\eqref{eqn:local-curvature} and, say, that which is obtained from numerical simulations must involve comparing two three-dimensional objects embedded in $\mathbb{R}^4$. Instead of attempting to do this, we have compared the approximation of the integrand of Eq.~\eqref{eqn:I-nw-general}, $\mathcal{G}/\rho_w^n$ (by substituting Eq.~\eqref{eqn:local-curvature}) with numerical values obtained from \texttt{Xsuite}~\cite{iadarola2023xsuite} for $n\in\{1,2,3\}$ along the $x$-, $y$- and $z$-axes for $G(x,s)=1$. The plots in the $x$-direction are reproduced here as an illustration. The upper part of Fig.~\ref{fig:comparisons} shows that for $n=1$, our analytical expressions well approximate the simulated behaviour; the $1\sigma$ cut-off of the approximated integrand is also fairly justifiable by the exponential suppression. Results for cuts along the $y$- and $z$-directions show similar good approximations for $n=1$; however, for higher values of $n$, our analytical formulae do not agree as well with the simulated values (see the middle and bottom part of Fig.~\ref{fig:comparisons} for the $x$-direction cuts). The behaviour of the integrand for cuts in the $y$- and $z$-directions show similar trends; notably, the analytical approximation overestimates the numerical values for the $y$-direction cuts (while underestimating it for cuts in the other two directions).

    General cases featuring first order approximations of the impact of the hourglass and of the crab waist are treated in App. B
    . For simplicity we focus here on the configuration neglecting both, i.e $G(s,x)_{w/s}\approx1$, so that we have,
	\begin{widetext}
		\begin{align} 
			\label{eqn:I-nhgncw-int}
			\mathcal{I}_{n,w}^{\text{\cancel{HG}, \cancel{CW}}}&\approx  \dfrac{1}{(2\pi)^{3/2}}\iiiint\displaylimits_{x,y,z,s\in\mathbb{R}^4}\dfrac{dxdydzds}{\sigma_{x,w}^*\sigma_{y,w}^*\sigma_{z,w}\sigma_{x,s}^{*n}}\left(\dfrac{2r_eN_s}{\gamma_w\sigma_{z,s}}\sqrt{\dfrac{2}{\pi}}\right)^n\left[\left(\dfrac{x-\left(s-{z}/{2}\right)\theta_c}{\sigma_{x,s}^*}\right)^2+\left(\dfrac{y}{\sigma_{y,s}^*}\right)^2\right]^{n/2}\nonumber\\
			&\times\exp\left[-\dfrac{n\left[x-\left(s-{z}/{2}\right)\theta_c\right]^2}{2\sigma_{x,s}^{*2}}-\dfrac{n(2s-z)^2}{2\sigma_{z,s}^2}-\dfrac{\left(x+{z\theta_c}/{2}\right)^2}{2\sigma_{x,w}^{*2}}-\dfrac{y^2}{2\sigma_{y,w}^{*2}}-\dfrac{z^2}{2\sigma_{z,w}^{2}}\right].
		\end{align}
	\end{widetext}
	The integral of interest was already evaluated analytically for symmetric bunch parameters with $n=1,3$ in Ref.~\cite{Garcia}. A more general derivation, which closely follows that in Ref.~\cite{Garcia} but is applicable for any $n$ as well as asymmetric bunch parameters, is given here. 
	
	Setting $x-\left(s-{z}/{2}\right)\theta_c=u$ as in Ref.~\cite{Garcia}, the negative of the exponent in the integrand can be rewritten in a form that facilitates Gaussian-like integrals,
	\begin{align}
		\label{eqn:exponent}
		n\dfrac{u^2}{2\sigma_{x,s}^{*2}}+n\dfrac{(2s-z)^2}{2\sigma_{z,s}^{*2}}+\dfrac{(u+s\theta_c)^2}{2\sigma_{x,w}^{*2}}+\dfrac{y^2}{2\sigma_{y,w}^{*2}}+\dfrac{z^2}{2\sigma_{z,w}^2}.
	\end{align}
	It is convenient to introduce the following terms
	\begin{align}
		\label{eqn:abcdeg}
		A&=\dfrac{n}{2\sigma_{x,s}^{*2}}+\dfrac{1}{2\sigma_{x,w}^{*2}},&B&=\dfrac{\theta_c}{\sigma_{x,w}^{*2}},&C&=\dfrac{2n}{\sigma_{z,s}^2}+\dfrac{\theta_c^2}{2\sigma_{x,w}^{*2}},\nonumber\\
		D&=\dfrac{n}{2\sigma_{z,s}^2}+\dfrac{1}{2\sigma_{z,w}^2},&E&=-\dfrac{2n}{\sigma_{z,s}^2},&G&=C-\dfrac{E^2}{4D},
	\end{align}
	such that the negative of the exponent can be re-written as
	\begin{align}
		G\left(s+\dfrac{Bu}{2G}\right)^2+\left(A-\dfrac{B^2}{4G}\right)u^2+D\left(z+\dfrac{Es}{2D}\right)^2+\dfrac{y^2}{2\sigma_{y,w}^{*2}}.
	\end{align}
	Performing the two Gaussian integrals over z and $s$ gives
	\begin{align}
		\label{eqn:int-nohgnocw}
		\mathcal{I}_{n,w}^{\text{\cancel{HG}, \cancel{CW}}}&=H_n\dfrac{\pi}{\sqrt{DG}}\iint\displaylimits_{u,y\in\mathbb{R}^2} dudy\left[\left(\dfrac{u}{\sigma_{x,s}^*}\right)^2+\left(\dfrac{y}{\sigma_{y,s}^*}\right)^2\right]^{n/2}\nonumber\\
		&\times\exp\left[-\left(A-\dfrac{B^2}{4G}\right)u^2-\dfrac{y^2}{2\sigma_{y,w}^{*2}}\right],
	\end{align}
	where
	\begin{equation}
		\label{eqn:Hn}
		H_n=\dfrac{1}{(2\pi)^{3/2}}\dfrac{1}{\sigma_{x,w}^*\sigma_{y,w}^*\sigma_{z,w}\sigma_{x,s}^{*n}}\left(\dfrac{2r_eN_s}{\gamma_s\sigma_{z,s}}\sqrt{\dfrac{2}{\pi}}\right)^n.
	\end{equation}
	Changing variables to $u/\sigma_{x,s}^*=r\cos\phi$ and $y/\sigma_{y,s}^*=r\sin\phi$ for $0<r<\infty$ and $0\leq\phi<\pi/2$, and performing the resultant Gaussian-like integral over $r$ using Ref.~\cite[Eq. (3.1.1)]{Wang&Guo} gives
	\begin{align}
		\mathcal{I}_{n,w}^{\text{\cancel{HG}, \cancel{CW}}}&=\dfrac{2H_n \Gamma\left(\frac{n}{2}+1\right)}{M_{n,w}^{\frac{n}{2}+1}}\dfrac{\pi}{\sqrt{DG}}\sigma_{x,s}^{*}\sigma_{y,s}^{*}\nonumber\\
		&\times\int\displaylimits_0^{\pi/2}d\phi\dfrac{1}{\left[1-P_{n,w}\sin^2\phi\right]^{\frac{n}{2}+1}},
	\end{align}
	in which we note that due to the symmetry in the range of $\phi$, a factor of 4 has been extracted to limit the range of integration over $\phi$ to $0\leq\phi<\pi/2$. We have also defined the new parameters
	\begin{align}
		M_{n,w}&=\left(A-\dfrac{B^2}{4G}\right)\sigma_{x,s}^{*2},&\Theta_w&=\dfrac{\sigma_{y,s}^*}{\sigma_{y,w}^*},&P_{n,w}&=1-\dfrac{\Theta_w^2}{2M_{n,w}}.
	\end{align}
	To perform the remaining integral over $\phi$, we change variables to $t=\sin^2\phi$ and define the constant
	\begin{equation}
		R_{n,w} = \pi\Gamma\left(\frac{n}{2}+1\right)H_n\frac{\pi}{\sqrt{DG}}\sigma_{x,s}^*\sigma_{y,s}^*
	\end{equation}
	to obtain
	\begin{equation}
		\label{eqn:I-nhgncw}
		\mathcal{I}_{n,w}^{\text{\cancel{HG}, \cancel{CW}}} = \dfrac{R_{n,w}}{M_{n,w}^{\frac{n}{2}+1}}F\left(\dfrac{n}{2}+1,\dfrac{1}{2},1;P_{n,w}\right),
	\end{equation}
	where the hypergeometric function $F(\alpha,\beta,\gamma;z)$ has the integral representation in Ref.~\cite[Eq.(4.5.6)]{Wang&Guo}. Expressed in terms of the beam quantities, the terms in Eq.~\eqref{eqn:I-nhgncw} read
	\begin{align}
		\label{eqn:mpr}
		M_{n,w}&=\dfrac{n}{2}+\dfrac{2n\sigma_{x,s}^{*2}}{\theta_c^2(n\sigma_{z,w}^2+\sigma_{z,s}^2)+4n\sigma_{x,w}^{*2}},\nonumber\\
		P_{n,w}&=1-\dfrac{({\sigma_{y,s}^{*2}})/({2\sigma_{y,w}^{*2}})}{\dfrac{n}{2}+\dfrac{2n\sigma_{x,s}^{*2}}{\theta_c^2(n\sigma_{z,w}^2+\sigma_{z,s}^2)+4n\sigma_{x,w}^{*2}}},\nonumber\\
		R_{n,w}&=\left(\dfrac{\pi}{2}\right)^{1/2}\Gamma\left(\dfrac{n}{2}+1\right)\dfrac{\sigma_{z,s}\sigma_{y,s}^*}{\sigma_{y,w}^*\sigma_{x,s}^{*n-1}}\left(\dfrac{2r_eN_s}{\gamma_w\sigma_{z,s}}\sqrt{\dfrac{2}{\pi}}\right)^n\nonumber\\
		&\hspace{20pt}\times\dfrac{1}{\sqrt{\theta_c^2(n\sigma_{z,w}^2+\sigma_{z,s}^2)+4n\sigma_{x,w}^{*2}}}.
	\end{align}
	A proof that this more general result reduces to those obtained for the special case of symmetric collisions as presented in Ref.~\cite{Garcia} is provided in App.~A
	.
	\subsection{Evaluating equilibrium bunch lengths using different approximations}
	\begin{table}
		\caption{\label{tab:FCC}Relevant FCC parameters, reproduced from Ref.~\cite{FCC}.}
		\begin{ruledtabular}
			\begin{tabular}{llccccc}
				{}&{}&\textbf{Z}&\textbf{WW}&\textbf{ZH}&\textbf{t\=t$_1$}&\textbf{t\=t$_2$}\\
				\hline
				$C$&[km]&\multicolumn{5}{c}{97.756}\\
				$\theta_c$&[mrad]&\multicolumn{5}{c}{30}\\
				Beam energy&[GeV]&45.6&80&120&175&182.5\\
				$N_0$&[$10^{11}$]&1.7&1.5&1.8&2.2&2.3\\
				$\alpha_p$&[$10^{-6}$]&{14.8}&{14.8}&{7.3}&{7.3}&{7.3}\\
				$\beta_x^*$&[m]&0.15&0.2&0.3&\multicolumn{2}{c}{1.0}\\
				$\beta_y^*$&[mm]&0.8&1.0&1.0&\multicolumn{2}{c}{1.6}\\
				$\sigma_{x}^*$&[$\mu$m]&6.4&13.0&13.7&36.7&38.2\\
				$\sigma_{y}^*$&[nm]&28&41&36&66&68\\
				$\sigma_{z,\text{SR}}$&[mm]&3.5&3.0&3.15&2.01&1.97\\
				$\Phi$&{}&28.5&7.0&5.8&1.1&1.0 \\
				$L_i$&[mm]&0.42&0.85&0.90&\multicolumn{2}{c}{1.8}\\
				$R_\text{HG}$&{}&0.95&0.89&0.88&\multicolumn{2}{c}{0.84}\\
				$Q_s$&{}&0.0250&0.0506&0.0358&0.0818&0.0872\\
				$\tau_{z,\text{SR}}$&[turns]&1273&236&70.3&23.1&20.4\\
    			$\tau_{z,\text{BS}}$&[turns]&186018&69011&26586&11239&10068\\
    			$U_{\text{SR}}$&[GeV]&0.036&0.34&1.72&7.8&9.2\\
    			$U_{\text{BS}}$&[MeV]&0.49&2.32&9.03&31.14&36.25\\
				$\xi_x/\xi_y$&{}&$\frac{0.004}{0.133}$&$\frac{0.010}{0.113}$&$\frac{0.016}{0.118}$&$\frac{0.097}{0.128}$&$\frac{0.099}{0.126}$\\
			\end{tabular}
		\end{ruledtabular}
	\end{table}

 	\begin{figure}\centering\includegraphics{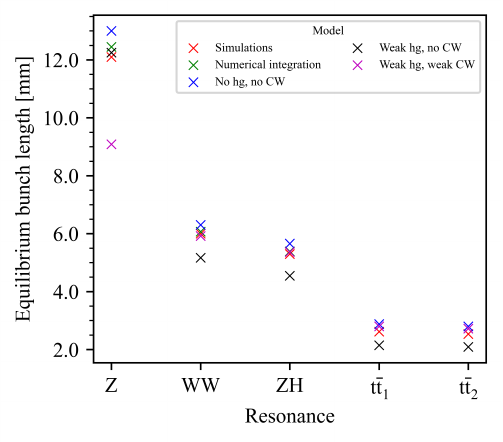}
		\caption{\label{fig:approx-vs-calc}Comparing the equilibrium bunch lengths, using FCC-ee parameters at the Z, WW, ZH and t\=t resonances (obtained from the various approximations, with hg being ``hourglass'' and CW being ``crab waist'') with those obtained from simulations in Ref.~\cite{Shatilov} and by integrating numerically Eq.~(B1) 
		as done in Ref.~\cite{Garcia} (for the latter, only data for Z, WW and ZH resonances are available).}
	\end{figure}
 
	For symmetric bunch parameters, the two equations of \eqref{eqn:evol_lengths} for the stationary case coincide. This is the case for the machine parameters proposed in the FCC-ee design study ~\cite{FCC}; relevant parameters are reproduced in Tab.~\ref{tab:FCC}. Fig.~\ref{fig:approx-vs-calc} shows the equilibrium bunch lengths obtained using these parameters and compares them to results previously published in Ref.~\cite{Shatilov} using simulations and in Ref.~\cite{Garcia} using direct numerical integration. For most configurations, the first order approximation of the hourglass and the crab waist (App. B2
	) quoted ``weak hg, weak CW'' yields the closest equilibrium bunch length to the simulations. This approximation is however rather poor for the Z configuration described in Tab.~\ref{tab:FCC}, as it features the strongest hourglass effect ($\sigma_z/\beta_y^*\sim10$). Due to the first order approximation, the correction terms in $\mathcal{O}[(\sigma_z/\beta_y^*)^2]$ (App. B
	) introduce strong artefacts in the integral. Taking aside this exception, the different approximations have a minor impact on the equilibrium bunch length and yield estimates comparable to simulations. The impact of the approximations was found to be minor also in the analysis of asymmetric configurations. Thus, in the following, we only show results using Eq.~\eqref{eqn:I-nhgncw} which neglects both the hourglass effect and the crab waist.
 
	\subsection{\label{subsec:lumi}Luminosity per interaction point at equilibrium}
 
\begin{figure}
		\centering
		\includegraphics[width = \linewidth]{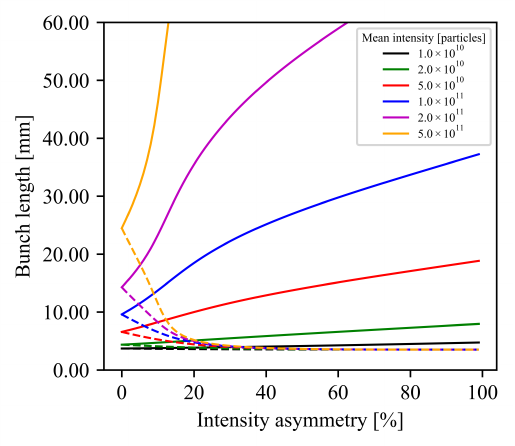}
		\caption{\label{paperfig:Z-length-1D}Equilibrium bunch lengths at the Z resonance determined with the outlined approximations using the first set of published machine parameters Tab.~\ref{tab:FCC} but with varying intensities of the two bunches. The solid line indicates the weak bunch length, while the dotted line gives that of the strong bunch. The FCC design study required both beams to have an intensity of $1.7\times10^{11}$ particles at this resonance.}
	\end{figure}

	We may now calculate the equilibrium length for the configurations featuring an asymmetry in the population of the two colliding bunches $\Delta N$ such that $N_{h/l}=N_0\times(1\pm\Delta N)$, with $N_h$, $N_l$ and $N_0$ the high, low and mean bunch intensity respectively. Though it is the most difficult configuration, due to the strong hourglass effect, we shall focus on the Z configuration of the FCC as this is where the tolerance on the asymmetry will turn out to be the tightest. The equilibirum bunch lengths of the two bunches are shown in Fig.~\ref{paperfig:Z-length-1D}. 
 
    At a given mean intensity, with increasing asymmetry, the weak bunch becomes longer without bound, while the strong bunch shortens to the length it would have if beamstrahlung were absent; both bunches are equally long when they have the same intensities, as would be expected from the symmetry of the system. The equilibrium bunch lengths for a given intensity asymmetry generally increase with the mean bunch intensity, as the effect of beamstrahlung becomes stronger, which is also in accordance with expectation. It should also be noted that as the mean intensity and asymmetry increase, at some point the bunch lengths obtained by these calculations ceases to be physical, as it exceeds the length of the longitudinal bucket. The corresponding beam losses are not taken into account in this simplistic model. 

\begin{figure}
		\includegraphics{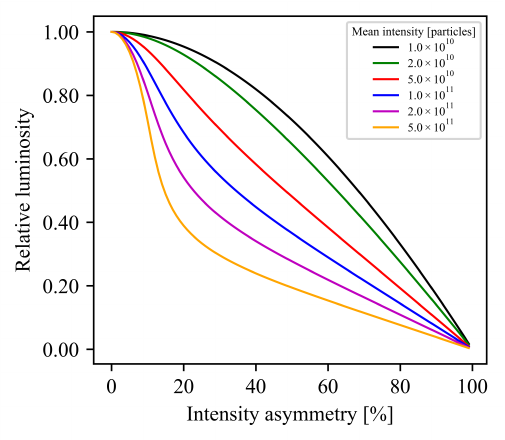}
		\caption{\label{paperfig:Z-luminosity-1D}Luminosity, for collisions at the Z resonance, normalised to the symmetric value at every mean intensity, determined with the outlined approximations using the corresponding published machine parameters in Tab.~\ref{tab:FCC} and the equilibrium lengths found via solving Eq.~\eqref{eqn:evol_lengths} for the stationary case.}
	\end{figure}

    For mean bunch intensities higher than $5\times10^{10}$ (red curve in Fig.~\ref{paperfig:Z-length-1D}) we observe a strong divergence of the bunch length of the low and high intensity bunch caused by the difference in the strength of beamstrahlung. This divergence results in a fast drop of the luminosity as a function of the bunch intensity asymmetry, as shown in Fig.~\ref{paperfig:Z-luminosity-1D}. 	
	
	The luminosity estimates assume again that corrections linked to the hourglass effect and the crab waist are negligible ~\cite{Damerau},	
	\begin{align}
		\label{eqn:lumi}
		\mathcal{L}&=\dfrac{N_lN_h}{T_\text{rev}}\dfrac{\cos(\theta_c/2)}{2\pi}\dfrac{1}{\sqrt{\sigma_{y,l}^{*2}+\sigma_{y,h}^{*2}}}\nonumber\\
		&\times\dfrac{1}{\sqrt{(\sigma_{x,l}^{*2}+\sigma_{x,h}^{*2})\cos^2(\theta_c/2)+(\sigma_{z,l}^2+\sigma_{z,h}^2)\sin^2(\theta_c/2)}}.
	\end{align}
    We have thus obtained a precursor of the 3D flip-flop mechanism by considering only the longitudinal effects linked to beamshtralung. In Sec.~\ref{sec:transverse}, we will address how the luminosity degradation is further enhanced by the transverse blow-up of the low intensity bunch due to the strong non-linearity of the beam-beam force generated by the short, high intensity bunch.

\subsection{\label{sec:stability}Stability of the stationary solutions}
	We now study the stability of the solution obtained with Eq.~\eqref{eqn:evol_lengths} in the stationary case. We define the right hand side of the equation as $f_{l/h}(\sigma_{z,l}^2,\sigma_{z,h}^2)$ and for simplicity we only consider the model without hourglass and crab waist. Using Ref.~\cite[Eq.(4.2.10)]{Wang&Guo} in the differentiation, we find,
	\begin{widetext}
		\begin{align}
			\label{eqn:fww}
			&\pdv{f_{l}}{\sigma_{z,l}^2} = \dfrac{2}{\tau_{z,\text{SR}}}\left\{\mathcal{A}_l\left[
			\left(\dfrac{M_{3,l}^{5/2}\pdv{R_{3,l}}{\sigma_{z,l}^2}-\dfrac{5}{2}R_{3,l}M_{3,l}^{3/2}\pdv{M_{3,l}}{\sigma_{z,l}^2}}{M_{3,l}^5}\right)F\left(\dfrac{5}{2},\dfrac{1}{2};1;P_{3,l}\right)+\dfrac{5}{4}\dfrac{R_{3,l}}{M_{3,l}^{5/2}}\pdv{P_{3,l}}{\sigma_{z,l}^2}F\left(\dfrac{7}{2},\dfrac{3}{2};2;P_{3,l}\right)\right]-1\right\}\nonumber\\
			&-\mathcal{B}_l\left\{\mathcal{I}_{2,l}+\sigma_{z,l}^2\left[\left(\dfrac{M_{2,l}^{2}\pdv{R_{2,l}}{\sigma_{z,l}^2}-2R_{2,l}M_{2,l}\pdv{M_{2,l}}{\sigma_{z,l}^2}}{M_{2,l}^4}\right)F\left(2,\dfrac{1}{2};1;P_{2,l}\right)+\dfrac{R_{2,l}}{M_{2,l}^2}\pdv{P_{2,l}}{\sigma_{z,l}^2}F\left(3,\dfrac{3}{2};2;P_{2,l}\right)\right]\right\}.\\
			\nonumber \\
			\label{eqn:fws}
			&\pdv{f_{l}}{\sigma_{z,h}^2} = \dfrac{2}{\tau_{z,\text{SR}}}\left\{\mathcal{A}_l\left[
			\left(\dfrac{M_{3,l}^{5/2}\pdv{R_{3,l}}{\sigma_{z,h}^2}-\dfrac{5}{2}R_{3,l}M_{3,l}^{3/2}\pdv{M_{3,l}}{\sigma_{z,h}^2}}{M_{3,l}^5}\right)F\left(\dfrac{5}{2},\dfrac{1}{2};1;P_{3,l}\right)+\dfrac{5}{4}\dfrac{R_{3,l}}{M_{3,l}^{5/2}}\pdv{P_{3,l}}{\sigma_{z,h}^2}F\left(\dfrac{7}{2},\dfrac{3}{2};2;P_{3,l}\right)\right]\right\}\nonumber\\
			&-\mathcal{B}_l\left\{\mathcal{I}_{2,l}+\sigma_{z,l}^2\left[\left(\dfrac{M_{2,l}^{2}\pdv{R_{2,l}}{\sigma_{z,s}^2}-2R_{2,l}M_{2,l}\pdv{M_{2,l}}{\sigma_{z,s}^2}}{M_{2,l}^4}\right)F\left(2,\dfrac{1}{2};1;P_{2,l}\right)+\dfrac{R_{2,l}}{M_{2,l}^2}\pdv{P_{2,l}}{\sigma_{z,h}^2}F\left(3,\dfrac{3}{2};2;P_{2,l}\right)\right]\right\},
		\end{align}
	\end{widetext}
	with the other two derivatives for $f_h$ given by exchanging indices in the above equations. We may evaluate the Jacobian at the equilibrium
	\begin{equation}
		J(\sigma_{z,l,\text{eqm}}^2,\sigma_{z,h,\text{eqm}}^2)=\left(\begin{matrix} \partial{f_l}/\partial{\sigma_{z,l}^2} & \partial{f_l}/\partial{\sigma_{z,h}^2} \\ \partial{f_h}/\partial{\sigma_{z,l}^2} & \partial{f_h}/\partial{\sigma_{z,h}^2} \end{matrix}\right)\Bigg|_\text{eqm}.
	\end{equation}
	It was found that, over a physical range of mean bunch intensities, at all intensity asymmetries, all resonances had two perturbation eigenvectors, $\bm{e}_{++}$ corresponding to the case where both bunches grow or shrink, and $\bm{e}_{+-}$ to the case where one bunch grows and the other one shrinks. The corresponding eigenvalues, $\lambda_{++}$ and $\lambda_{+-}$, were found to always be negative. The sign of the eigenvalues implies that all equilibrium configurations of the bunch lengths are stable to perturbation. One reason for this is due to the damping effect of beamstrahlung: while the beam-beam interaction is necessary for the blow-up to occur, as the beam intensity increases, at higher bunch intensities and asymmetries, where beamstrahlung damping dominates the damping effect of synchrotron radiation emitted from bending in the arcs, the weak bunch is more strongly shortened by the emitted radiation than it is lengthened by the resulting energy spread, and therefore any lengthening effect due to the field of the strong bunch is quickly dissipated away. 
	
\begin{figure}
		\subfloat[\label{paperfig:++eig-Z}Z, $\lambda_{++}$]{
			\includegraphics{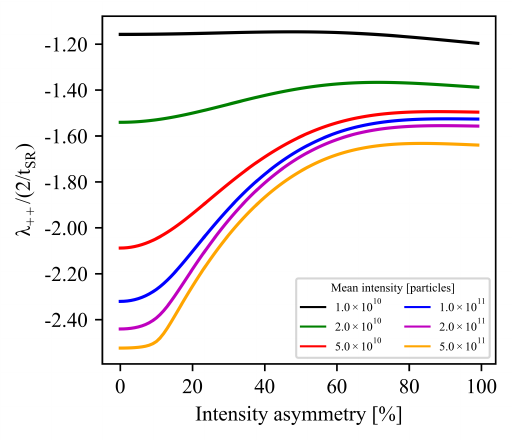}
		}
		\newline	
		\subfloat[\label{paperfig:+-eig-Z}Z, $\lambda_{+-}$]{
			\includegraphics{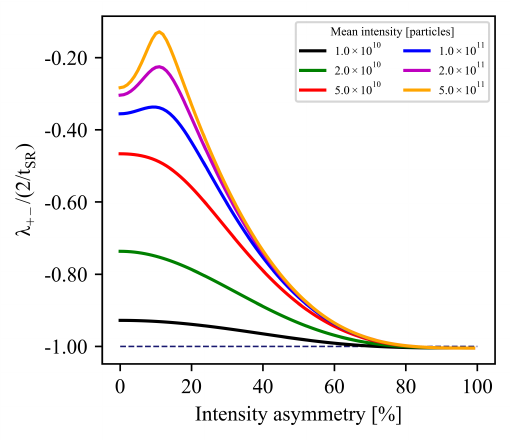}
		}
		\caption{\label{paperfig:eig-Z}Eigenvalues obtained in performing linear stability analysis on the equilibrium bunch lengths at different mean bunch intensities and intensity asymmetry of collisions at the Z resonance in the ``no hourglass, no crab waist'' approximation. They are normalised to the decay rate of the emittance due to synchrotron radiation $2/\tau_\text{{SR}}$, indicated by the broken line on Fig.~\ref{paperfig:+-eig-Z}. It can be seen that as the intensity asymmetry approaches 100\%, beamsstrahlung is no longer active, and the one of the eigenvalues approaches this synchrotron-induced decay rate.}
	\end{figure}

	Nevertheless, the study of the eigenvalue coresponding to an asymmetric perturbation of the bunch lengths $\lambda_{+-}$ in Fig.~\ref{paperfig:+-eig-Z} shows regions where the perturbations are expected to decay on time scales longer than the one expected from the radiation damping in the arc only. On the other hand, Fig.~\ref{paperfig:++eig-Z} shows that the decay time corresponding to a symmetric perturbation can be significantly faster.
	
	Consequently, one may expect that the transient perturbation due to the top-up injection of additional particles to the bunches could decay on longer time scale. For the configuration studied here, namely the FCC-ee at the Z pole, the time scale remains short enough not to affect the luminosity significantly.
	\section{Transverse blow-up}\label{sec:transverse}
	\begin{figure}
		\includegraphics[width = \linewidth]{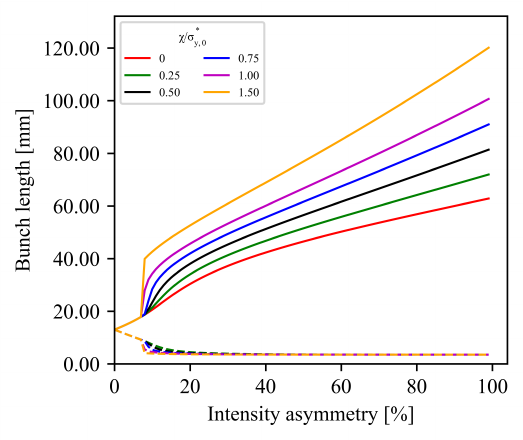}
		\caption{\label{paperfig:Z-length-3D}Equilibrium bunch lengths at the Z resonance using the 3D phenomenological model with different $\chi/\sigma_{y,0}^*$. The solid line indicates the weak bunch's length, while the dotted line gives that of the strong bunch. The red line, showing no $zy$-coupling, is identical to that which would have been obtained by applying the 1D ``no hourglass, no crab waist'' model to the published machine parameters in Tab.~\ref{tab:FCC} for a symmetric collision. Here, $\xi_0 = 0.20.$}
	\end{figure}
	\begin{figure}
		\includegraphics{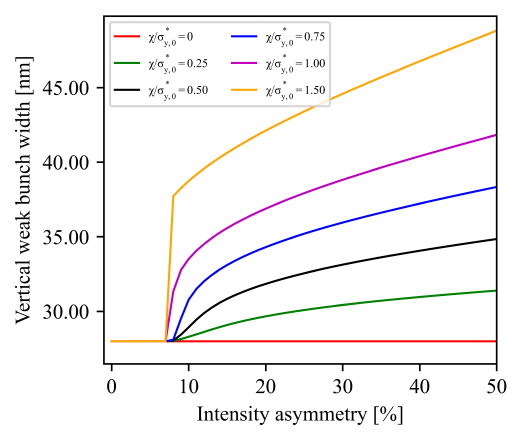}
		\caption{\label{paperfig:Z-width-3D} Equilibrium vertical width of the weak bunch at the Z resonance using different approximations in the 3D phenomenological model with different $\chi/\sigma_{y,0}^*$. The range of intensity asymmetries has been truncated to show more clearly the behaviour of the weak bunch's width in the two regimes. Here, $\xi_0 = 0.20.$}
	\end{figure}
	The 3D flip-flop phenomenon as discussed in Ref.~\cite{Shatilov} involves an increase of the transverse dimensions of the low intensity bunch caused by non-linear diffusion driven by the beam-beam force of the higher intensity (and shorter) bunch, and coupling between the longitudinal and transverse beam-beam forces due to the collision with a large crossing angle. In quantifying this effect, the beam-beam parameters $\xi$ is often used, which, in the case of flat bunches ($\sigma_y\ll\sigma_x$) with a non-zero crossing angle $\theta_c$ and large Piwinski angle $\Phi\gg1$, take the form~\cite{Shatilov},
	\begin{align}
		\xi_{x,w}(\sigma_{z,s})&\approx\dfrac{N_{s}r_e}{\pi\gamma_{s}}\times\dfrac{2\beta_x^*}{(\sigma_{z,s}\theta_c)^2},\\
		\label{eqn:xi-yw}
		\xi_{y,w}(\sigma_{z,s}, \sigma_{y,s}^*)&\approx\dfrac{N_{s}r_e}{\pi\gamma_{s}}\times\dfrac{\beta_y^*}{\sigma_{z,s}\sigma_{y,s}^*\theta_c}.
	\end{align}
	The flat beam geometry usually implies that $\xi_{y}>\xi_{x}$. Moreover, since $N_l<N_h$ by assumption and $\sigma_{z,l}<\sigma_{z,h}$ as previously found, the vertical beam-beam parameter of the low intensity bunch $\xi_{y,l}$ is the largest. Therefore, we shall coarsely model a transverse-longitudinal flip-flop, with the blowup limited to the y-z plane, with a phenomenological model of the vertical size of the low intensity bunch parameterised by $\xi_{y,l}$. We assume that the vertical blow up can be neglected for a beam-beam strength below a given threshold $\xi_0$. Above this threshold the vertical beam size increases proportionally to the excess of beam-beam parameters:
	\begin{equation} \label{eqn:y-blowup}
    \sigma_{y,l,\text{new}}^* =
        \left\lbrace
        \begin{array}{ll}
        \sigma_{y,l,0}^*&\text{if}~\xi < \xi_0 \\
        \sigma_{y,l,0}^*+\chi\Delta\xi_y&\text{otherwise},
        \end{array}
        \right.
    \end{equation}
	with $\chi$ the phenomenological transversal blow-up factor and $\Delta\xi_y$ the excess of vertical beam-beam tune shift:
 \begin{equation}
     \Delta\xi_y = \xi_{y,l,\text{new}}(\sigma_{z,h,\text{eqm, new}}, \sigma_{y,h}^* = \sigma_{y,0}^*) - \xi_0.
 \end{equation}
 The nonlinear dependence of the beam sizes on $\xi_{x,y}$ is key to the onset of the transverse-longitudinal flip-flop effect. Note that Eq.~\ref{eqn:y-blowup} is nonlinear in the sense that it is a threshold function, representing regular motion for weak enough non-linear forces (\emph{i.e.} $\xi<\xi_0$) with an onset of transverse diffusion at a given $\xi_0$. Clearly, the behavior much beyond the threshold is more nonlinear than the linear dependence assumed by the model, yet its aim is to describe the dynamics close to the threshold $\xi_0$.The phenomenological model has indeed no predictive power for strong beam-beam strength ($\xi \gg \xi_0$).
 
	Considering the $\xi_y$ values in Tab.~\ref{tab:FCC}, we chose $\xi_0=0.2$ as a reasonable threshold beam interaction value which is not much higher than that for symmetric collisions at the machine parameters. The threshold beam-beam parameter defines the boundary between two different regimes: the $zy$-decoupled (1D) and the $zy$-coupled (2D) regimes. The transition between these regimes is visible in Figs.~\ref{paperfig:Z-length-3D} and ~\ref{paperfig:Z-width-3D}. In the 2D regime, the introduction of coupling between the transverse and longitudinal dimensions of the weak bunch to the original two equilibrium equations causes the blow up in equilibrium bunch length and bunch width to occur more rapidly with increasing asymmetry in the bunch intensities and increasing $\chi$. This corresponds to a simplified model of the 3D flip-flop described in Ref.~\cite{Shatilov}, without coupling to the horizontal plane. We note however that this phenomenon is different in nature to the transverse flip-flop observed in lower energies electron-positron colliders featuring negligible beamstrahlung~\cite{Tennyson}. Indeed, here the dynamical system does not feature multiple equilibrium solutions. Nevertheless, the runaway of the low intensity bunch parameters towards lowering the beam-beam force and of the high intensity bunch towards increasing the beam-beam force in the transverse-longitudinal flip-flop results in a sensitivity to the asymmetry between the two beams which is comparable to the purely transverse flip-flop phenomenon. 
	
\begin{figure}
		\centering
		\includegraphics{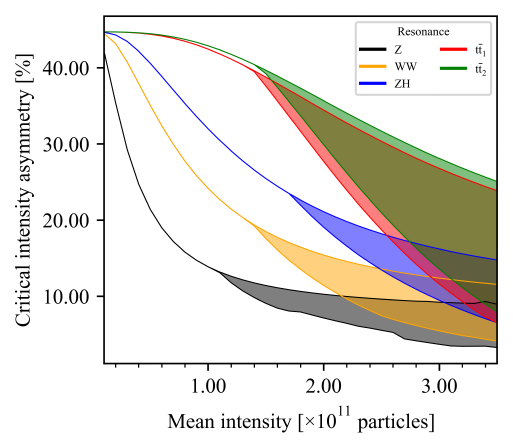}
		\caption{Critical intensity asymmetry for which the relative intensity does not drop below 80\% using the 3D ``no hourglass, no crab waist'' model. The shaded region for every resonance represents values of $\chi/\sigma_{y,0}^*$ lying between 0 (the upper bound, for which there is no coupling between the z- and y-dimensions and so reducing to the 1D model) and 1.5 (the lower bound). Here, $\xi_0 = 0.20.$}
		\label{paperfig:critical-asymmetry}
	\end{figure}

	Fig.~\ref{paperfig:critical-asymmetry} shows the tolerance for intensity asymmetry such that the luminosity remains above 80\% of that of a completely symmetric collision. The most sensitive configuration of the FCC-ee corresponds to the Z pole: where an asymmetry of only 8\% at $\chi=1.5\sigma_{y,0}^*$ can be tolerated. This tolerance is comparable to the 5\% quoted for the FCC-ee design~\cite{FCC}, yet we note that the phenomenological parameters $\chi$ and $\xi_0$ used here were coarsely estimated. Numerical simulations without approximating the beam-beam force and the approximated phenomenological model for the transverse size increase due to non-linear diffusion are still needed to determine suitable values for the phenomenological parameters. Given that they are driven by the non-linear dynamics of the particles, those parameters are expected to strongly depend on the working point chosen.
	\section{Conclusion}
	Integrals that give the average energy loss and quantum excitation due to beamstrahlung effects in two crossing Gaussian bunches of arbitrary bunch dimensions were generalised to configurations featuring asymmetric machine and beam parameters. The hourglass effect and the crab waist were taken into account to first order, but their impact on the equilibrium beam parameters remained marginal. 
	
	Thanks to these integrals, the impact of the intensity asymmetry on the equilibrium bunch lengths of the two beams could be studied. The low intensity bunch tends to increase in length due to the increase in beamstrahlung caused by the high intensity bunch. On the other hand, the high intensity bunch tends to shrink towards the value determined by the magnetic lattice. The equilibrium solutions were found to be stable. However in some configurations an initial asymmetric perturbation, \emph{e.g.} during top-up injection, could damp with a rate significantly lower than the one given by synchrotron radiation damping. 
	
	A plausible phenomenological model, which includes transverse blow up of the weak bunch's vertical width due to non-linear diffusion, was proposed as a way to model the vertical blow-up observed in simulations. Though arguably coarse, the phenomenological model yields a tolerance on the intensity asymmetry comparable to those obtained in past studies with simulations.
	
	\acknowledgements{We thank T. v. Witzleben for a careful reading of the manuscript in its early stages. This work was carried out in the framework of an undergraduate student summer project as part of the Excellence Research Internship Programme hosted by the \'Ecole Polytechnique F\'ed\'erale de Lausanne (EPFL), under an exchange agreement with the University of Cambridge, at the European Organisation for Nuclear Research (CERN) and under the auspices of and with support from the Swiss Accelerator Research and Technology (CHART) programme (\url{www.chart.ch}).}

	\bibliography{bibliography}

\end{document}


\title{Supplemental material to the manuscript\\
		"Impact of bunch intensity asymmetry in colliders featuring strong beamstrahlung"}
	\maketitle
	\appendix
	\section{Verifying $\mathcal{I}_{n,w}$ for negligible hourglass and crab waist effects and symmetric bunch parameters}\label{app:proof-Garcia}
	In the paper, the following expression, Eq.~(23)
	, was obtained in the limit of no hourglass and crab waist effect for the integral $\mathcal{I}_{n,w}$,
	\begin{equation}
		\label{appeqn:I-nw-nhgncw}
		\mathcal{I}_{n,w}^{\cancel{\text{HG}}\cancel{\text{CW}}} =  \frac{R_{n,w}}{M_{n,w}^{ \frac{n}{2}+1}}F\left( \frac{n}{2}+1, \frac{1}{2},1;P_{n,w}\right),
	\end{equation}
	where 	
	\begin{align}
		M_{n,w}&=\dfrac{n}{2}+\dfrac{2n\sigma_{x,s}^{*2}}{\theta_c^2(n\sigma_{z,w}^2+\sigma_{z,s}^2)+4n\sigma_{x,w}^{*2}},\\
		P_{n,w}&=1-\dfrac{\dfrac{\sigma_{y,s}^{*2}}{2\sigma_{y,w}^{*2}}}{\dfrac{n}{2}+\dfrac{2n\sigma_{x,s}^{*2}}{\theta_c^2(n\sigma_{z,w}^2+\sigma_{z,s}^2)+4n\sigma_{x,w}^{*2}}},\\
		R_{n,w}&=\left(\dfrac{\pi}{2}\right)^{1/2}\Gamma\left(\dfrac{n}{2}+1\right)\dfrac{\sigma_{z,s}\sigma_{y,s}^*}{\sigma_{y,w}^*\sigma_{x,s}^{*n-1}}\left(\dfrac{2r_eN_s}{\gamma_w\sigma_{z,s}}\sqrt{\dfrac{2}{\pi}}\right)^n\times\dfrac{1}{\sqrt{\theta_c^2(n\sigma_{z,w}^2+\sigma_{z,s}^2)+4n\sigma_{x,w}^{*2}}}.
	\end{align}
	We will show that expressions derived in Ref.~\cite{Garcia} for the case of collisions between symmetric bunches are obtained as appropriate special cases of Eq.~\eqref{appeqn:I-nw-nhgncw}. In this case, the subscripts $s$ and $w$ are unimportant (as quantities with these two subscripts are equal, for example $\sigma_{x,s}^*=\sigma_{x,w}^*\equiv\sigma_x^*$). Defining the Piwinski angle $\Phi\equiv({\theta_c\sigma_z})/({2\sigma_x^*})$, it can be shown that the above parameters simplify to
	\begin{align}
		\label{appeqn:MnPnRn}
		M_n^\text{sym}&= \frac{n(n+1)(1+\Phi^2)}{2[n+(n+1)\Phi^2]},\nonumber\\
		P_n^\text{sym}&=1- \frac{n+(n+1)\Phi^2}{n(n+1)(1+\Phi^2)},\nonumber\\
		R_n^\text{sym}&=\frac{1}{2}\left(\frac{\pi}{2}\right)^{1/2}\Gamma\left( \frac{n}{2}+1\right) \frac{1}{\sigma_{z}^{n-1}}\left( \frac{2r_eN}{\gamma\sigma_x^*}\sqrt{\frac{2}{\pi}}\right)^n \frac{1}{\sqrt{n+(n+1)\Phi^2}}.
	\end{align}
	Defining $\mathcal{Q}_n$ as in Ref.~\cite{Garcia} as
	\begin{align}
		\mathcal{Q}_n&\equiv\frac{1}{\sigma_z^{n-1}}\left(\frac{2Nr_e}{\gamma\sigma_x^*}\sqrt{\frac{2}{\pi}}\right)^n,
	\end{align}
	we see that
	\begin{equation}
		R_n^\text{sym}=\frac{1}{2}\left(\frac{\pi}{2}\right)^{1/2}\Gamma\left(\frac{n}{2}+1\right)\frac{\mathcal{Q}_n}{\sqrt{n+(n+1)\Phi^2}}.
	\end{equation}
	Firstly, we show that Eq.~\eqref{appeqn:I-nw-nhgncw} coincides with the expression for $n=1$ given in Ref.~\cite{Garcia} for the case of symmetric bunches colliding with no crossing angle, such that $\Phi=0$. Then, $M_1^\text{sym}=1$, $P_1^\text{sym}=1/2$, $R_1^\text{sym}=(\pi \mathcal{Q}_1)/(4\sqrt{2})$, using $\Gamma(3/2)=\sqrt{\pi}/2$, and hence
	\begin{equation}
		\label{appeqn:I-1}
		\mathcal{I}_1^\text{sym, \cancel{HG}, \cancel{CW}, head-on} = \frac{\pi \mathcal{Q}_1}{4\sqrt{2}}F\left(\frac{3}{2},\frac{1}{2};1;\frac{1}{2}\right).
	\end{equation}
	The hypergeometric function $F(\alpha,\beta,\gamma;z)$ satisfies the following relationship with its ``close neighbours" $F(\alpha\pm1,\beta,\gamma;z)$ \cite[Eq.~(4.2.7)]{Wang&Guo}
	\begin{equation}
		\label{appeqn:neighbours}
		[\gamma-2\alpha+(\alpha-\beta)z]F(\alpha)+\alpha(1-z)F(\alpha+1)-(\gamma-\alpha)F(\alpha-1)=0,
	\end{equation}
	where only transformations on $\alpha$ are indicated. Applying this relationship to $\alpha=1/2$, $\beta=1/2$, $\gamma=1$ and $z=1/2$ gives
	\begin{equation}
		\label{appeqn:F-1-equation}
		\frac{1}{4}F\left(\frac{3}{2},\frac{1}{2},1;\frac{1}{2}\right)-\frac{1}{2}F\left(-\frac{1}{2},\frac{1}{2};1;\frac{1}{2}\right)=0.
	\end{equation}
	The complete elliptic integral of the second kind, $E(k^2)$, is defined as
	\begin{equation}
		E(k^2)\equiv\int_0^1\sqrt{ \frac{1-k^2t^2}{1-t^2}}dt,
	\end{equation}
	and is related to the hypergeometric function when $k<1$ as \cite{Wang&Guo}
	\begin{equation}
		\label{appeqn:elliptic-second}
		E(k^2)=\frac{\pi}{2}F\left(-\frac{1}{2},\frac{1}{2},1;k^2\right).
	\end{equation}
	Eqs. \eqref{appeqn:F-1-equation} and \eqref{appeqn:elliptic-second}, when taken together, gives
	\begin{equation}
		\label{appeqn:E-half}
		F\left(\frac{3}{2},\frac{1}{2},1;\frac{1}{2}\right)=\frac{4}{\pi}E\left(\frac{1}{2}\right).
	\end{equation}
	Combining Eq.~\eqref{appeqn:I-1} with Eq.~\eqref{appeqn:E-half} yields the result published in Ref.~\cite{Garcia},
	\begin{equation}
		\mathcal{I}_1^\text{sym, \cancel{HG}, \cancel{CW}, head-on} = \frac{\mathcal{Q}_1}{\sqrt{2}}E\left(\frac{1}{2}\right).
	\end{equation}
	Next, we demonstrate that this expression yields the same result derived in Ref.~\cite{Garcia} for $n=3$ in the case where the bunches have symmetric parameters. 
	For the case $n=3$, defining $\varphi$ as in Ref.~\cite{Garcia} as
	\begin{equation}
		\varphi\equiv \frac{4\Phi^2}{3+4\Phi^2},
	\end{equation}
	we obtain
	\begin{align}
		\label{appeqn:mpr-Garcia}
		M_3^\text{sym}&= \frac{6(1+\Phi^2)}{3+4\Phi^2},\nonumber\\
		P_3^\text{sym}&= \frac{9+8\Phi^2}{12(1+\Phi^2)}= \frac{3-\varphi}{4-\varphi},\nonumber\\
		R_3^\text{sym}&= \frac{3\pi}{8\sqrt{2}} \frac{\mathcal{Q}_3}{\sqrt{3+4\Phi^2}},
	\end{align}
	where we have used $\Gamma\left( \frac{5}{2}\right)= \frac{3}{4}\sqrt{\pi}$. 
	
	Applying Eq.~\eqref{appeqn:neighbours} twice with $\alpha, \beta, \gamma=(3/2,1/2,1)$ and $(1/2,1/2,1)$, we can show that
	\begin{equation}
		\frac{3}{2}(1-z)F\left( \frac{5}{2}, \frac{1}{2},1;z\right)= \left(\frac{2-z}{1-z}\right)F\left( \frac{-1}{2}, \frac{1}{2},1;z\right)+ \frac{1}{2}F\left( \frac{1}{2}, \frac{1}{2},1;z\right),
	\end{equation}
	Hence, for $z=P_3^\text{sym}=(3-\varphi)/(4-\varphi)$, it can be further shown that
	\begin{equation}
		F\left(\frac{5}{2}, \frac{1}{2},1; \frac{3-\varphi}{4-\varphi}\right) =  \frac{2}{3}(5-\varphi)(4-\varphi)F\left(- \frac{1}{2}, \frac{1}{2},1; \frac{3-\varphi}{4-\varphi}\right)+ \frac{1}{3}(4-\varphi)F\left( \frac{1}{2}, \frac{1}{2},1; \frac{3-\varphi}{4-\varphi}\right).
	\end{equation}
	On the other hand, the complete elliptic integrals of the first kind, $K(k^2)$, with the following definition,
	\begin{equation}
		K(k^2)\equiv\int_0^1 \frac{dt}{\sqrt{(1-t^2)(1-k^2t^2)}},\nonumber\\
	\end{equation}
	is related to the hypergeometric function when $k<1$ as \cite{Wang&Guo}
	\begin{equation}
		K(k^2)= \frac{\pi}{2}F\left(\frac{1}{2}, \frac{1}{2},1;k^2\right),\nonumber\\
	\end{equation}
	Therefore, it follows that
	\begin{equation}
		\label{appeqn:ek-Garcia}
		F\left( \frac{5}{2}, \frac{1}{2},1; \frac{3-\varphi}{4-\varphi}\right)= \frac{1}{2\pi}\left( \frac{4-\varphi}{3}\right)\left[(40-8\varphi)E\left( \frac{3-\varphi}{4-\varphi}\right)-4K\left( \frac{3-\varphi}{4-\varphi}\right)\right].
	\end{equation}
	Combining Eqs.~\eqref{appeqn:mpr-Garcia} and \eqref{appeqn:ek-Garcia}, after some algebra, leads to
	\begin{equation}
		\mathcal{I}_{3}^\text{sym, \cancel{HG}, \cancel{CW}} = \mathcal{Q}_3 \frac{\left[(10-2\varphi)E\left( \frac{3-\varphi}{4-\varphi}\right)-K\left( \frac{3-\varphi}{4-\varphi}\right)\right]}{\sqrt{(3+4\Phi^2)(4-\varphi)^3}},
	\end{equation}
	which is the same result derived in Ref.~\cite{Garcia}.
	\section{Approximations to the integral $\mathcal{I}_{n,w}$} \label{app:approx}
	In this section, we will derive the approximations to the integral,
	\begin{align} \label{appeqn:full-int}
		\mathcal{I}_{n,w}&=   \frac{1}{(2\pi)^{3/2}}\iiiint\displaylimits_{x,y,z,s\in\mathbb{R}^4} \frac{dxdydzds}{\sigma_{x,w}^*\sigma_{y,w}^*\sigma_{z,w}\sigma_{x,s}^{*n}}\left( \frac{2r_eN_s}{\gamma_w\sigma_{z,s}}\sqrt{ \frac{2}{\pi}}\right)^n\left[1+\left( \frac{s-x/\theta_c}{\beta_y^*}\right)^2\right]^{-1/2}\nonumber\\
		&\times\left[\left( \frac{x-\left(s-z/2\right)\theta_c}{\sigma_{x,s}^*}\right)^2+\left( \frac{y}{\sigma_{y,s}^*}\right)^2\left[1+\left( \frac{s+x/\theta_c}{\beta_y^*}\right)^2\right]^{-1}\right]^{n/2}\nonumber\\
		&\times\exp\left[- \frac{n\left[x-\left(s-z/2\right)\theta_c\right]^2}{2\sigma_{x,s}^{*2}}- \frac{n(2s-z)^2}{2\sigma_{z,s}^2}- \frac{\left(x+{z\theta_c}/{2}\right)^2}{2\sigma_{x,w}^{*2}}- \frac{y^2}{2\sigma_{y,w}^{*2}}\left[1+\left( \frac{s-x/\theta_c}{\beta_y^*}\right)^2\right]^{-1}- \frac{z^2}{2\sigma_{z,w}^{*2}}\right],
	\end{align}
	in the case of weak hourglass effect and no crab waist effect, and in the case of weak hourglass effect and weak crab waist effect, as used in the paper.
	\subsection{Weak hourglass, negligible crab waist effect}
	\label{app:weak_hg_no_CW}
	If the contribution of the crab waist scheme to the integrand is negligible, and the hourglass effect is weak for $s\sim L_i$, then we can rewrite the factor $G(s,x)_w$ approximately as $G(s,x)_w=\sqrt{1+[(s\pm x/\theta_c)/\beta_y^*]^2}\approx\sqrt{1+(s/\beta_y^*)^2}$ and expand Eq.~\eqref{appeqn:full-int} as a Taylor series in $(s/\beta_y^*)^2$, keeping only terms $O[(s/\beta_y^*)^2]$. Combining this with the change of variables $u=x-(s-z/2)\theta_c$ and rewriting the exponent of the integrand as done in Eq.~(15)
	 leads to
	\begin{align} 
		&\mathcal{I}_{n,w}^\text{weak HG, \cancel{CW}}\approx \frac{1}{(2\pi)^{3/2}}\iiiint\displaylimits_{x,y,z,s\in\mathbb{R}^4} \frac{dxdydzds}{\sigma_{x,w}^*\sigma_{y,w}^*\sigma_{z,w}\sigma_{x,s}^{*n}}\left( \frac{2r_eN_s}{\gamma_w\sigma_{z,s}}\sqrt{ \frac{2}{\pi}}\right)^n\left[1- \frac{1}{2}\left( \frac{s}{\beta_y^*}\right)^2\right]\nonumber\\
		&\times\left[\left( \frac{u}{\sigma_{x,s}^*}\right)^2+\left( \frac{y}{\sigma_{y,s}^*}\right)^2\left[1-\left( \frac{s}{\beta_y^*}\right)^2\right]\right]^{n/2}\exp\left[-G\left(s+ \frac{Bu}{2G}\right)^2-\left(A- \frac{B^2}{4G}\right)u^2-D\left(z+ \frac{Es}{2D}\right)^2- \frac{y^2}{2\sigma_{y,w}^{*2}}\left[1-\left( \frac{s}{\beta_y^*}\right)^2\right]\right].
	\end{align}
	After performing the integral over $z$, collecting the constants and further expanding $\exp[({y^2}/{2\sigma_{y,w}^{*2}})({s}/{\beta_y^*})^2]$ as a Taylor series to $O[(s/\beta_y^*)^2]$, we obtain
	\begin{align}
		\label{eqn:before-s-expand-weakhgnocw}
		\mathcal{I}_{n,w}^\text{weak HG, \cancel{CW}}\approx H_n\sqrt{ \frac{\pi}{D}}\iiint\displaylimits_{u,y,s\in\mathbb{R}^3} dudyds&\left[1- \frac{1}{2}\left( \frac{s}{\beta_y^*}\right)^2\right]\left[1-\left( \frac{n}{2}\right) \frac{\left({y}/{\sigma_{y,s}^*}\right)^2}{\left( {u}/{\sigma_{x,s}^*}\right)^2+\left({y}/{\sigma_{y,s}^*}\right)^2}\left( \frac{s}{\beta_y^*}\right)^2\right]\left[1+ \frac{y^2}{2\sigma_{y,w}^{*2}}\left( \frac{s}{\beta_y^*}\right)^2\right]\nonumber\\
		&\times\left[\left( \frac{u}{\sigma_{x,s}^*}\right)^2+\left( \frac{y}{\sigma_{y,w}^*}\right)^2\right]^{n/2}\exp\left[G-\left(s+ \frac{Bu}{2G}\right)^2-\left(A- \frac{B^2}{4G}\right)u^2- \frac{y^2}{2\sigma_{y,w}^{*2}}\right],
	\end{align}
	where $H_n$ is defined as in Eq.~(19)
	. The terms containing $(s/\beta_y^*)^2$ in the first line of Eq.~\eqref{eqn:before-s-expand-weakhgnocw} can further be expanded to $O[(s/\beta_y^*)^2]$, which gives
	\begin{align}
		\label{eqn:beforesplit-nohgweakcw}
		\mathcal{I}_{n,w}^\text{weak HG, \cancel{CW}}\approx H_n\sqrt{ \frac{\pi}{D}}\iiint\displaylimits_{u,y,s\in\mathbb{R}^3} dudyds &\left[1- \frac{1}{2}\left( \frac{s}{\beta_y^*}\right)^2\left[1+n\left[ \frac{\left( \frac{y}{\sigma_{y,s}^*}\right)^2}{\left( \frac{u}{\sigma_{x,s}^*}\right)^2+\left( \frac{y}{\sigma_{y,s}^*}\right)^2}\right]-\left( \frac{y}{\sigma_{y,w}^*}\right)^2\right]\right]\nonumber\\
		&\times\left[\left( \frac{u}{\sigma_{x,s}^*}\right)^2+\left( \frac{y}{\sigma_{y,s}^*}\right)^2\right]^{n/2}\exp\left[-G\left(s+ \frac{Bu}{2G}\right)^2-\left(A- \frac{B^2}{4G}\right)u^2- \frac{y^2}{2\sigma_{y,w}^{*2}}\right].
	\end{align}
	Comparing with Eq.~(18)
	, we see that the two terms in the first line of Eq.~\eqref{eqn:beforesplit-nohgweakcw} split into $\mathcal{I}_{n,w}^\text{\cancel{HG}, \cancel{CW}}$ and a correction term, which we shall assume to be small,
	\begin{align}
		\mathcal{I}_{n,w}^\text{weak HG, \cancel{CW}}\approx\mathcal{I}_{n,w}^\text{\cancel{HG}, \cancel{CW}}- \frac{1}{2}H_n\sqrt{ \frac{\pi}{D}}\iiint\displaylimits_{u,y,s\in\mathbb{R}^3} dudyds &\left( \frac{s}{\beta_y^*}\right)^2\left[1+n\left[ \frac{\left({y}/{\sigma_{y,s}^*}\right)^2}{\left({u}/{\sigma_{x,s}^*}\right)^2+\left({y}/{\sigma_{y,s}^*}\right)^2}\right]-\left( \frac{y}{\sigma_{y,w}^*}\right)\right]\left[\left( \frac{u}{\sigma_{x,s}^*}\right)^2+\left( \frac{y}{\sigma_{y,s}^*}\right)^2\right]^{n/2}\nonumber\\
		&\times\exp\left[-G\left(s+ \frac{Bu}{2G}\right)^2-\left(A- \frac{B^2}{4G}\right)u^2- \frac{y^2}{2\sigma_{y,w}^{*2}}\right].
	\end{align}
	We will need the result
	\begin{equation}
		\label{eqn:gaussian-2}
		\int\displaylimits_{-\infty}^{+\infty} (a+bx)^2e^{-c(x+d)^2}dx=\dfrac{b^2\sqrt{\pi}}{2c^{3/2}}\left[\dfrac{2c}{b^2}(a-bd)^2+1\right],
	\end{equation}
	valid for $c>0$. The proof is simple:
	\begin{align}
		\int\displaylimits_{-\infty}^{+\infty} (a+bx)^2e^{-c(x+d)^2}dx&=\int\displaylimits_{-\infty}^{+\infty}[a+b(x-d)]^2e^{-cx^2}dx\nonumber\\
		&=\int\displaylimits_{-\infty}^{+\infty}[(a-bd)^2+2abx+b^2x^2]e^{-cx^2}dx\nonumber\\
		&=\dfrac{b^2\sqrt{\pi}}{2c^{3/2}}\left[\dfrac{2c}{b^2}(a-bd)^2+1\right];
	\end{align}
	the second term in the integrand is odd and integrates to zero, while the other two terms are evaluated using Ref.~\cite[Eq.~(3.1.1)]{Wang&Guo}. Therefore, performing the integration over $s$ using Eq.~\eqref{eqn:gaussian-2} gives
	\begin{align}
		\mathcal{I}_{n,w}^\text{weak HG, \cancel{CW}}\approx\mathcal{I}_{n,w}^\text{\cancel{HG}, \cancel{CW}} -  \frac{1}{2}H_n\sqrt{ \frac{\pi}{D}} \frac{1}{2\beta_y^{*2}}\iint\displaylimits_{u,y\in\mathbb{R}^2} dudy& \frac{\sqrt{\pi}\left( \frac{B^2u^2}{2G}+1\right)}{2G^{3/2}}\left[1+n\left[ \frac{\left( {y}/{\sigma_{y,s}^*}\right)^2}{\left({u}/{\sigma_{x,s}^*}\right)^2+\left({y}/{\sigma_{y,s}^*}\right)^2}\right]-\left( \frac{y}{\sigma_{y,w}^*}\right)^2\right]\nonumber\\
		&\times\left[\left( \frac{u}{\sigma_{x,s}^*}\right)^2+\left( \frac{y}{\sigma_{y,s}^*}\right)^2\right]^{n/2}\exp\left[-\left(A- \frac{B^2}{4G}\right)u^2- \frac{y^2}{2\sigma_{y,w}^{*2}}\right].
	\end{align}
	Changing variables to $u/\sigma_{x,s}^* = r\cos\phi$ and $y/\sigma_{y,s}^*=r\sin\phi$ and performing some algebra gives
	\begin{align}
		\mathcal{I}_{n,w}^\text{weak HG, \cancel{CW}}=\mathcal{I}_{n,w}^\text{\cancel{HG}, \cancel{CW}} -  \frac{R_{n,w}}{4\pi G\beta_y^{*2}\Gamma(\frac{n}{2}+1)}\int\displaylimits_{r=0}^\infty\int\displaylimits_{\phi=0}^{2\pi}&drd\phi r^{n+1}\left(1+ \frac{B^2\sigma_{x,s}^{*2}r^2\cos^2\phi}{2G}\right)\nonumber\\
		&\times[1+(n-\Theta_{w}^2)\sin^2\phi]\exp[-M_{n,w}r^2(1-P_{n,w}\sin^2\phi)],
	\end{align}
	where the parameters $R_{n,w}$, $M_{n,w}$ and $\Theta_{w}$ have been defined in the paper. Using Ref.~\cite[Eq.~(3.1.1)]{Wang&Guo} to perform the integrals over $r$ and the identity in Ref.~\cite[Eq.~(3.2.1)]{Wang&Guo}, $\Gamma(z+1)=z\Gamma(z)$, to rewrite the resultant gamma functions in terms of $\Gamma(n/2+1)$, and then changing variables to $v=\sin^2\phi$ (noting that the symmetry in the integrand with respect to $\phi\in[0,2\pi)$ leads to an extra factor of 4 due to this variable change), gives
	\begin{align}
		\mathcal{I}_{n,w}^\text{weak HG, \cancel{CW}}=\mathcal{I}_{n,w}^\text{\cancel{HG}, \cancel{CW}}-&\frac{R_{n,w}}{4\pi G\beta_y^{*2}M_{n,w}^{n/2+1}}\int_0^1 dv v^{-1/2}(1-v)^{-1/2}[1-(\Theta_w^2-n)v]\nonumber\\
		&\times\left[(1-P_{n,w}v)^{-(n/2+1)}+\frac{B^2\sigma_{x,s}^{*2}}{2GM}\left(\frac{n}{2}+1\right)(1-v)(1-P_{n,w}v)^{-(n/2+2)}\right].
	\end{align}
	We use the integral representation of the Appell $F_1$ function \cite[Eq.~(4.16.14)]{Wang&Guo} to evaluate the integral over $v$ and obtain
	\begin{align}\label{eqn:I-whgncw}
		\mathcal{I}_{n,w}^\text{weak HG, \cancel{CW}}&=\mathcal{I}_{n,w}^\text{\cancel{HG}, \cancel{CW}}\nonumber\\
		&-\frac{R_n}{2G\beta_y^{*2}M^{\frac{n}{2}+1}}\times\left[F_1\left(\frac{1}{2};-1,\frac{n}{2}+1;1;\Theta_w^2-n,P_{n,w}\right)+\frac{B^2\sigma_{x,s}^{*2}}{4GM_{n,w}}\left(\frac{n}{2}+1\right)F_1\left(\frac{1}{2};-1,\frac{n}{2}+2;2;\Theta_w^2-n,P_{n,w}\right)\right].
	\end{align}
	\subsection{\label{app:weak_hg_weak_CW}Weak hourglass and weak crab waist}
	We can perform a similar approximation if we assume that in the range for which the integrand has significant values, $(s\pm x/\theta_c)/\beta_y^*\ll1$. If this is the case, then following a similar method as in Appendix \ref{app:weak_hg_no_CW} and keeping terms up to $O[((s\pm x/\theta_c)/\beta_y^*)^2]$, we obtain the approximation
	\begin{align}
		\mathcal{I}_{n,w}^\text{weak HG, weak CW}\approx H_n\iiiint\displaylimits_{u,y,z,s\in\mathbb{R}^4}&dudydzds\left[\left(\frac{u}{\sigma_{x,s}^*}\right)^2+\left(\frac{y}{\sigma_{y,s}^*}\right)^2\right]^{n/2}\exp\left[-G\left(s+\frac{Bu}{2G}\right)^2-\left(A-\frac{B^2}{4G}\right)u^2-D\left(z+\frac{Es}{2D}\right)^2-\frac{y^2}{2\sigma_{y,w}^{*2}}\right]\nonumber\\
		&\times\left[1-\frac{1}{2}\left(\frac{\frac{z}{2}-\frac{u}{\theta_c}}{\beta_y^*}\right)\left[1-\left(\frac{y}{\sigma_{y,w}^*}\right)^2\right]-\frac{n}{2}\left[\frac{\left({y}/{\sigma_{y,s}^*}\right)^2}{\left({u}/{\sigma_{x,s}^*}\right)^2+\left({y}/{\sigma_{y,s}^*}\right)^2}\right]\left(\frac{s+u/\theta_c}{\beta_y^*}\right)^2\right],
	\end{align}
	which again breaks up into $\mathcal{I}_{n,w}^\text{\cancel{HG}, \cancel{CW}}$ and two correction terms, which we shall assume to both be small,
	\begin{align}
		\mathcal{I}_{n,w}^\text{weak HG, weak CW}&\approx\mathcal{I}_{n,w}^\text{\cancel{HG}, \cancel{CW}}&\nonumber\\
		&\hspace{-20pt}-H_n\iiiint\displaylimits_{u,y,z,s\in\mathbb{R}^4} dudydzds&\left[\left(\frac{u}{\sigma_{x,s}^*}\right)^2+\left(\frac{y}{\sigma_{y,s}^*}\right)^2\right]^{n/2}\exp\left[-G\left(s+\frac{Bu}{2G}\right)^2-\left(A-\frac{B^2}{4G}\right)u^2-D\left(z+\frac{Es}{2D}\right)^2-\frac{y^2}{2\sigma_{y,w}^{*2}}\right]\nonumber\\
		&{}&\times\frac{1}{2\beta_y^{*2}}\left(\frac{z}{2}-\frac{u}{\theta_c}\right)^2\left[1-\left(\frac{y}{\sigma_{y,w}^*}\right)^2\right]\nonumber\\
		&\hspace{-20pt}-H_n\iiiint\displaylimits_{u,y,z,s\in\mathbb{R}^4} dudydzds &\left[\left(\frac{u}{\sigma_{x,s}^*}\right)^2+\left(\frac{y}{\sigma_{y,s}^*}\right)^2\right]^{n/2}\exp\left[G-\left(s+\frac{Bu}{2G}\right)^2-\left(A-\frac{B^2}{4G}\right)u^2-D\left(z+\frac{Es}{2D}\right)^2-\frac{y^2}{2\sigma_{y,w}^{*2}}\right]\nonumber\\
		&{}&\times\frac{n}{2\beta_y^{*2}}\left[\frac{\left({y}/{\sigma_{y,s}^{*}}\right)^2}{\left({u}/{\sigma_{x,s}^*}\right)^2+\left({y}/{\sigma_{y,s}^*}\right)^2}\right]\left[s+u/\theta_c\right]^2.
	\end{align}
	We consider the first correction term, $\Circled{1}$. Using Eq.~\eqref{eqn:gaussian-2}, we can evaluate the integrals over $z$ and then over $s$ to obtain
	\begin{align}
		\Circled{1}=-\frac{H_n\sqrt{\pi}}{16D^{3/2}\beta_y^{*2}}\iint\displaylimits_{u,y\in\mathbb{R}^2} dudy&\left[\left(\frac{u}{\sigma_{x,s}^*}\right)^2+\left(\frac{y}{\sigma_{y,s}^*}\right)^2\right]^{n/2}\exp\left[-\left(A-\frac{B^2}{4G}\right)u^2-\frac{y^2}{2\sigma_{y,w}^{*2}}\right]\nonumber\\
		&\times\left\{2D\frac{\left(\frac{E}{2D}\right)^2\sqrt{\pi}}{2G^{3/2}}\left[\frac{2G}{\left({E}/{2D}\right)^2}\left(\frac{2u}{\theta_c}-\frac{E}{2D}\frac{Bu}{2G}\right)^2+1\right]+\sqrt{\frac{\pi}{G}}\right\}\left[1-\left(\frac{y}{\sigma_{y,w}^*}\right)^2\right].
	\end{align}
	We proceed exactly as previously by first changing variables to $u/\sigma_{x,s}^*=r\cos\phi$ and $y/\sigma_{y,s}^*=r\sin\phi$. Then, we use Ref.~\cite[Eq.~(3.1.1)]{Wang&Guo} to perform the Gaussian-like integrals over $r$ and Ref.~\cite[Eq.~(3.2.1)]{Wang&Guo} to rewrite the resultant gamma functions in terms of $\Gamma(n/2+1)$. We then change variables to $v=\sin^2\phi$, noting the quadrupling due to the symmetry in the range of $v$, and use Ref.~\cite[Eq.~(4.16.14)]{Wang&Guo} to evaluate the integral over $v$. The first correction term then works out to be
	\begin{align}
		\Circled{1}&=-\frac{R_{n,w}}{16DM_{n,w}^{\frac{n}{2}+1}\beta_y^{*2}}\left[\left(\frac{n}{2}+1\right)\frac{\sigma_{x,s}^{*2}D}{M_{n,w}}\left(\frac{2}{\theta_c}-\frac{EB}{4DG}\right)^2F_1\left(\frac{1}{2};-1,\frac{n}{2}+2;2;\Theta_w^2,P_{n,w}\right)+\left(1+\frac{E^2}{4DG}\right)F_1\left(\frac{1}{2};-1,\frac{n}{2}+1;1;\Theta_w^2,P_{n,w}\right)\right],
	\end{align}
	which, using the explicit definitions of the parameters $B$, $D$, $E$ and $G$ in Eq.~(16)
	, can be shown to be equal to
	\begin{equation}
		\Circled{1}=-\frac{R_{n,w}}{16M_{n,w}^{\frac{n}{2}+1}\beta_y^{*2}}\left[\mathcal{C}_1F_1\left(\frac{1}{2};-1,\frac{n}{2}+2;2;\Theta_w^2,P_{n,w}\right)+\mathcal{C}_2F_1\left(\frac{1}{2};-1,\frac{n}{2}+1;1;\Theta_w^2,P_{n,w}\right)\right],
	\end{equation}
	with the parameters $\mathcal{C}_{1,2}$ defined in Eq.~\eqref{eqn:whgwcw-parameters}. The second correction term, \Circled{2}, can be evaluated using exactly the same steps, the only difference being that Ref.~\cite[eq.~(4.6.6)]{Wang&Guo}, instead of Ref.~\cite[Eq.~(4.16.14)]{Wang&Guo} , is required to handle the hypergeometric, instead of Appell, functions that result from the integral over $v$. The result can be shown to be
	\begin{equation}
		\Circled{2}=-\frac{nR_{n,w}}{16\beta_y^{*2}GM_{n,w}^{\frac{n}{2}+1}}\left[\left(\frac{n}{2}+1\right)\frac{G\sigma_{x,s}^{*2}}{M_{n,w}}\left(\frac{1}{\theta_c}-\frac{B}{2G}\right)^2F\left(\frac{n}{2}+2,\frac{3}{2},3;P_{n,w}\right)+2F\left(\frac{n}{2}+1,\frac{3}{2},2;P_{n,w}\right)\right]
	\end{equation}
	which can be rewritten as
	\begin{equation}
		\Circled{2}=-\frac{nR_{n,w}}{16\beta_y^{*2}M_{n,w}^{\frac{n}{2}+1}}\left[\mathcal{D}_1F\left(\frac{n}{2}+2,\frac{3}{2},3;P_{n,w}\right)+\mathcal{D}_2F\left(\frac{n}{2}+1,\frac{3}{2},2;P_{n,w}\right)\right],
	\end{equation}
	using expressions for the parameters $B$, $D$, $E$ and $G$ in Eq.~(16) 
	to arrive at the parameters $\mathcal{D}_{1,2}$ as defined in Eq.~\eqref{eqn:whgwcw-parameters}. Therefore, the integral including weak hourglass and crab waist effects is approximately, \emph{i.e.}
	\begin{align}  \label{eqn:I-whgwcw}
		\mathcal{I}_{n,w}^\text{weak HG, weak CW}\approx\mathcal{I}_{n,w}^\text{\cancel{HG}, \cancel{CW}}-\frac{R_{n,w}}{16\beta_y^{*2}M_{n,w}^{\frac{n}{2}+1}}&\Bigg\{\left[\mathcal{C}_1F_1\left(\frac{1}{2};-1,\frac{n}{2}+2;2;\Theta_w^2,P_{n,w}\right)+\mathcal{C}_2F_1\left(\frac{1}{2};-1,\frac{n}{2}+1;1;\Theta_w^2,P_{n,w}\right)\right]\nonumber\\
		&+n\left[\mathcal{D}_1F\left(\frac{n}{2}+2,\frac{3}{2},3;P_{n,w}\right)+\mathcal{D}_2F\left(\frac{n}{2}+1,\frac{3}{2},2;P_{n,w}\right)\right]\Bigg\},
	\end{align}
	where 
	\begin{align}
		\label{eqn:whgwcw-parameters}
		\mathcal{C}_1&\equiv\left(\dfrac{n}{2}+1\right)\dfrac{\sigma_{x,s}^{*2}}{M_{n,w}}\left(\dfrac{2}{\theta_c}+\dfrac{2n\theta_c\sigma_{z,w}^2}{\sigma_{x,s}^{*2}}\Lambda_{n,w}^2\right)^2,\nonumber\\
		\mathcal{C}_2&\equiv\frac{1}{D}\left[1+4n^2\dfrac{\sigma_{z,w}^2\sigma_{x,w}^{*2}}{\sigma_{z,s}^2\sigma_{x,s}^{*2}}\Lambda_{n,w}^2\right],\nonumber\\
		\mathcal{D}_1&\equiv\left(\dfrac{n}{2}+1\right)\dfrac{\sigma_{x,s}^{*2}}{M_{n,w}}\left[\frac{1}{\theta_c}-\dfrac{\theta_c}{\sigma_{x,s}^{*2}}(n\sigma_{z,w}^2+\sigma_{z,s}^2)\Lambda_{n,w}^2\right]^2,\nonumber\\
		\mathcal{D}_2&\equiv\frac{2}{G}.
	\end{align}
	\section{\label{app:Jacobian}The Jacobian in linear stability analysis of equilibrium bunch lengths in the ``no hourglass, no crab waist'' model}
	In this Appendix, we state expressions for the remaining unevaluated partial derivatives in Eqs.~(26)  
	and (27) 
	for the elements $\partial f_w/\partial\sigma_{z,w}^2$ and $\partial f_w/\partial\sigma_{z,s}^2$ of the Jacobian at an equilibrium configuration determined via the ``no hourglass, no crab waist'' model; the other partial derivatives relevant to $f_s$ can be found by appropriately exchanging the indices $w$ and $s$. 
	
	The dimensionless quantities
	\begin{align}
		\label{eqn:lambda-q}
		\Lambda_{n,w}&\equiv\dfrac{\sigma_{x,s}^{*}}{\sqrt{\theta_c^2(n\sigma_{z,w}^2+\sigma_{z,s}^2)+4n\sigma_{x,w}^{*2}}},&\qquad
		Q_{n,w}&\equiv\left(\frac{\pi}{2}\right)^{1/2}\Gamma\left(\frac{n}{2}+1\right)\dfrac{1}{\sigma_{z,s}^{n-1}}\left(\dfrac{2r_eN_s}{\gamma_w\sigma_{x,s}^{*}}\sqrt{\dfrac{2}{\pi}}\right)^n,
	\end{align}
	can be used to write the parameters of Eq.~(24) 
	compactly as
	\begin{alignat}{3}
		\label{appeqn:mpr_short}
		M_{n,w}=\frac{n}{2}+2n\Lambda_{n,w}^2,&\qquad P_{n,w}=1-\frac{\Theta_w^2}{2M_{n,w}},&\qquad R_{n,w}=Q_{n,w}\Theta_w\Lambda_{n,w}.
	\end{alignat}
	
	First, we consider the partial derivatives of $\Lambda_{n,w}$ and $Q_{n,w}$, defined in Eq.~\eqref{eqn:lambda-q}, with respect to $\sigma_{z,w/s}^2$,
	\begin{alignat}{4}
		\label{appeqn:jacobian-pdv-1}
		\pdv{\Lambda_{n,w}}{\sigma_{z,w}^2}&=-\frac{n\Lambda_{n,w}^3\theta_c^2}{2\sigma_{x,s}^{*2}},&\qquad \pdv{\Lambda_{n,w}}{\sigma_{z,s}^2}&=-\frac{\Lambda_{n,w}^3\theta_c^2}{2\sigma_{x,s}^{*2}},&\qquad
		\pdv{Q_{n,w}}{\sigma_{z,w}^2}&=0,&\qquad
		\pdv{Q_{n,w}}{\sigma_{z,s}^2}&=-\frac{(n-1)Q_{n,w}}{2\sigma_{z,s}^2}.	
	\end{alignat}
	The product rule applied to Eq.~\eqref{appeqn:mpr_short} then simply gives 
	\begin{alignat}{3}
		\label{appeqn:jacobian-pdv-2}
		\pdv{M_{n,w}}{\sigma_{z,w}^2}&=4n\Lambda_{n,w}\pdv{\Lambda_{n,w}}{\sigma_{z,w}^2},&\qquad\pdv{P_{n,w}}{\sigma_{z,w}^2}&=\frac{\Theta_w^2}{2M_{n,w}^2}\pdv{M_{n,w}}{\sigma_{z,w}^2},&\qquad\pdv{R_{n,w}}{\sigma_{z,w}^2}&=Q_{n,w}\Theta_w\pdv{\Lambda_{n,w}}{\sigma_{z,w}^2},\nonumber\\
		\pdv{M_{n,w}}{\sigma_{z,s}^2}&=4n\Lambda_{n,w}\pdv{\Lambda_{n,w}}{\sigma_{z,s}^2},&\qquad\pdv{P_{n,w}}{\sigma_{z,s}^2}&=\frac{\Theta_w^2}{2M_{n,w}^2}\pdv{M_{n,w}}{\sigma_{z,s}^2},&\qquad\pdv{R_{n,w}}{\sigma_{z,s}^2}&=\Theta_w\left(\Lambda_{n,w}\pdv{Q_{n,w}}{\sigma_{z,s}^2}+Q_{n,w}\pdv{\Lambda_{n,w}}{\sigma_{z,s}^2}\right).
	\end{alignat}
	Equations \eqref{appeqn:jacobian-pdv-1} and \eqref{appeqn:jacobian-pdv-2} together allow for evaluations of Eqs.~(26) 
	and (27)
	.
	\section{\label{app:eqm-length}Equilibrium bunch lengths at different mean intensities for different intensity asymmetries for all resonances using different models with no transverse coupling}	
	Fig.~\ref{appfig:bunch-length-1d} shows the equilibrium bunch lengths determined by applying the various approximations to the 1D model with no transverse coupling between the length and the width of the weak bunch, \emph{i.e.} by solving Eq.~(11) 
	at equilibrium. For a given mean bunch intensity, the solid and broken lines indicate the lengths of the weaker and stronger bunches respectively. Mean intensities are varied to $\pm2\times10^{10}$ in steps of $1\times10^{10}$ around the published design mean intensity in Table I
	; the red line shows the result for this parameter.
	\section{\label{app:lumi}Luminosities at different mean intensities for different intensity asymmetries for all resonances using different models with no transverse coupling}
\begin{figure}[!h]
		\centering
		\subfloat[Z]{
			\includegraphics[width=.475\linewidth]{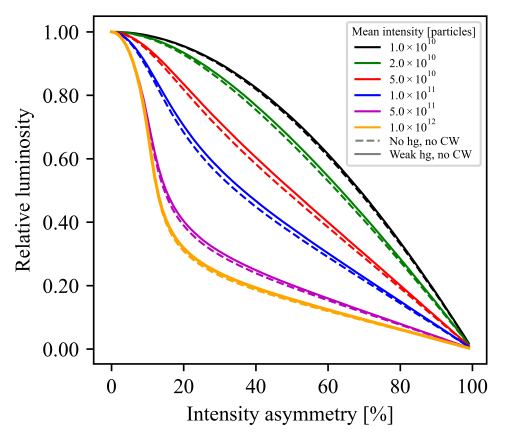}
		}
		\newline
		\subfloat[WW]{
			\includegraphics[width=.475\linewidth]{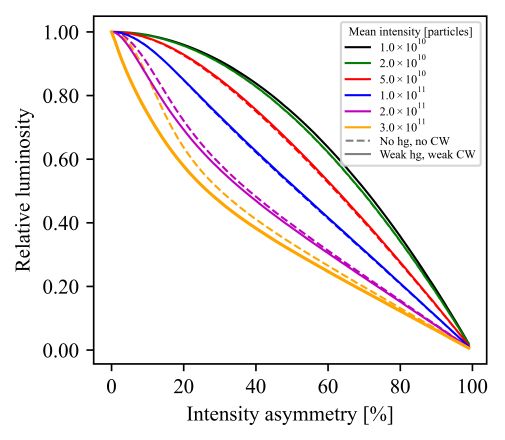}
		}
		\subfloat[ZH]{
			\includegraphics[width=.475\linewidth]{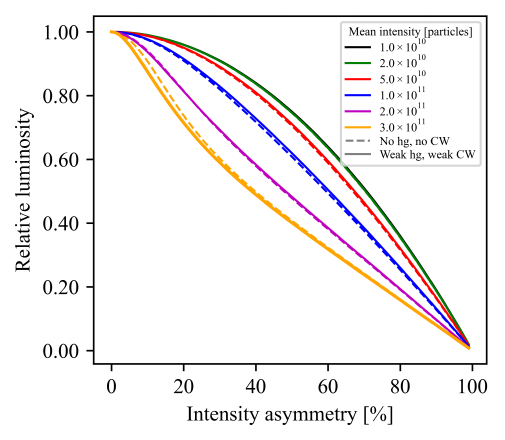}
		}
		\newline
		\subfloat[First t\=t]{
			\includegraphics[width=.475\linewidth]{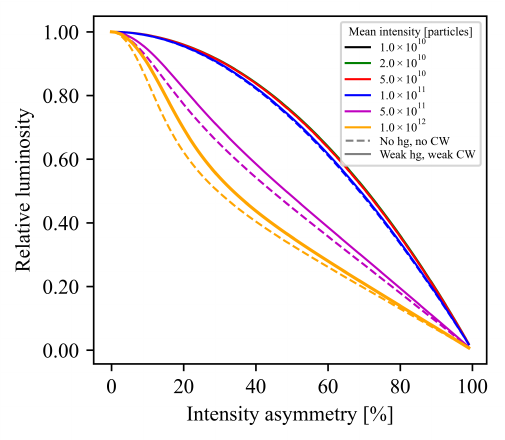}
		}
		\subfloat[Second t\=t]{
			\includegraphics[width=.475\linewidth]{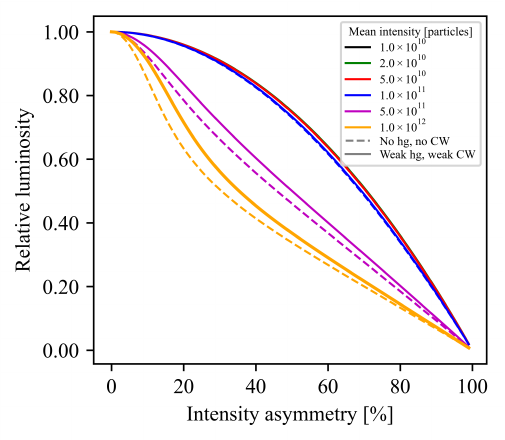}
		}
		\caption{\label{appfig:lumi-1d}Luminosity, normalised to the value at zero intensity asymmetry, for various mean intensities and intensity asymmetry of the various resonances using different approximations to the 1D model. Other parameters are taken from Tab.~I
		.}
	\end{figure}

	Similarly, Fig.~\ref{appfig:lumi-1d} shows the luminosities at different resonances determined by applying different approximations (with or without ``weak hourglass'', and with or without ``crab waist'') to the 1D model. The luminosities at every mean bunch intensity and for every asymmetry between the individual intensities relative to the mean was calculated using Eq.~(25) 
	and then normalised to the value at zero asymmetry. The mean bunch intensity was plotted over a range of an order of magnitude around the proposed intensity in table I
	, while the asymmetry was allowed to vary from 0\% to 99\%.
	\section{\label{app:eig}Eigenvalues at different mean intensities for different intensity asymmetries for all resonances in the ``no hourglass, no crab waist'' model with no transverse coupling}
	The linear stability analysis done in section II E 
	in the paper was only applicable to the 1D ``no hourglass, no crab waist'' model, so Fig.~\ref{appfig:eig-1D-nohgnocw} shows the behaviour of the two eigenvalues $\lambda_{++}$ and $\lambda_{+-}$, normalised to $2/\tau_\text{{SR}}$, as deduced from this model.
	\section{\label{app:3d-length-width}Equilibrium bunch lengths and weak bunch width using coupled models for different phenomenological factors of $\chi$ at $\xi_0=0.2$.}
\begin{figure}[!h]
		\centering
		\subfloat[Z, $\sigma_{y,w,\text{eqm}}$]{
			\includegraphics[width=.475\linewidth]{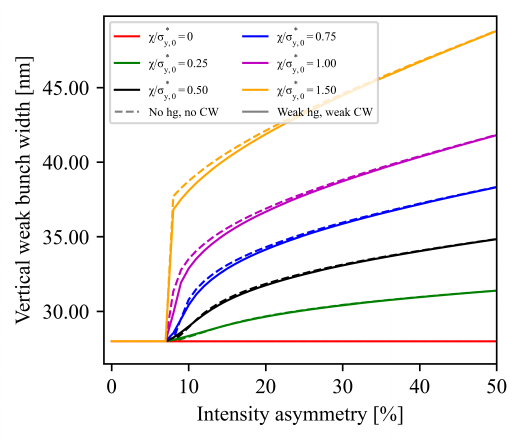}
		}
		\newline
		\subfloat[WW, $\sigma_{y,w,\text{eqm}}$]{
			\includegraphics[width=.475\linewidth]{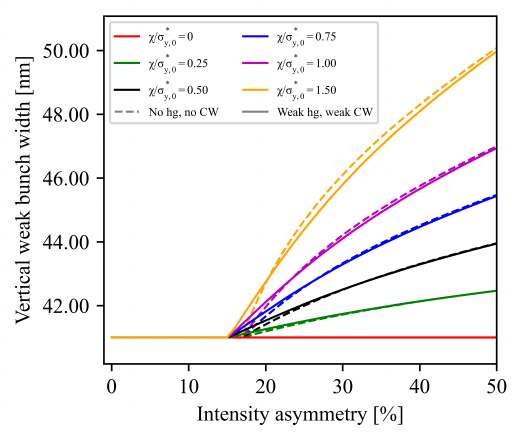}
		}
		\subfloat[ZH, $\sigma_{y,w,\text{eqm}}$]{
			\includegraphics[width=.475\linewidth]{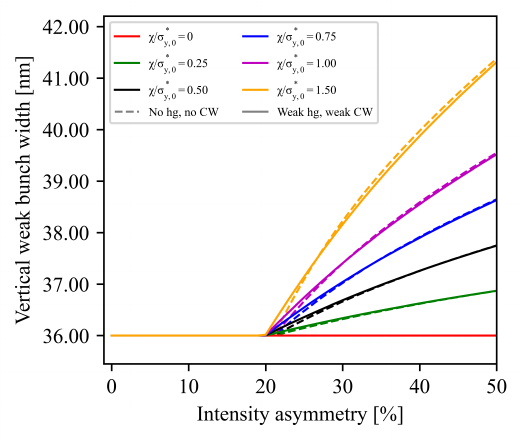}
		}
		\newline
		\subfloat[First t\=t, $\sigma_{y,w,\text{eqm}}$]{
			\includegraphics[width=.475\linewidth]{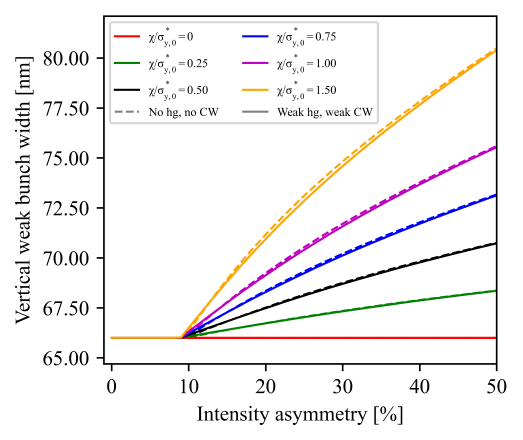}
		}
		\subfloat[Second t\=t, $\sigma_{y,w,\text{eqm}}$]{
			\includegraphics[width=.475\linewidth]{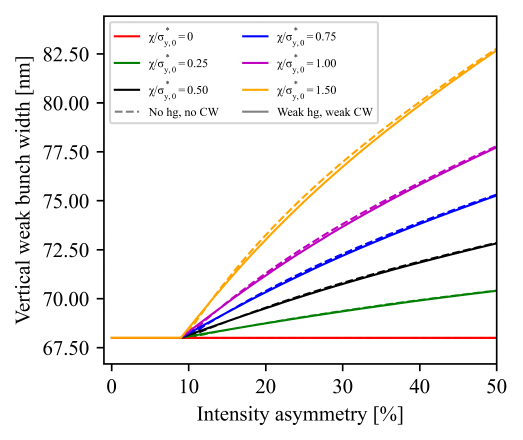}
		}
		\caption{\label{appfig:3d-length-width-weakhgweakcw}Equilibrium bunch lengths and vertical weak bunch width determined using machine parameters and the 3D ``weak hourglass, weak crab waist'' approximation at the various resonances, except the Z resonance, for which the 3D ``weak hourglass, no crab waist'' approximation was used. The threshold beam-beam parameter was fixed at $\xi_0=0.2$. The plots were done with machine parameters, including the mean intensity, as found in Tab.~I
		. For the equilibrium bunch lengths, the solid line shows the weak bunch length, the dotted line indicates that of the strong bunch and the red line, where no coupling was imposed, gives the 1D result.}
	\end{figure}

	Figs.~\ref{appfig:3d-length-width-nohgnocw} and \ref{appfig:3d-length-width-weakhgweakcw} show the equilibrium bunch lengths and weak bunch widths using different models, as indicated in the caption, where coupling between the bunches' transversal and longitudinal dimensions based on the proposed phenomenological model was allowed. The threshold beam-beam parameter was fixed at $\xi_0 = 0.2$.
	\pagebreak
	\begin{figure}[!h]
		\centering
		\subfloat[Z, no hourglass, no crab waist]{
			\includegraphics[width=.475\linewidth]{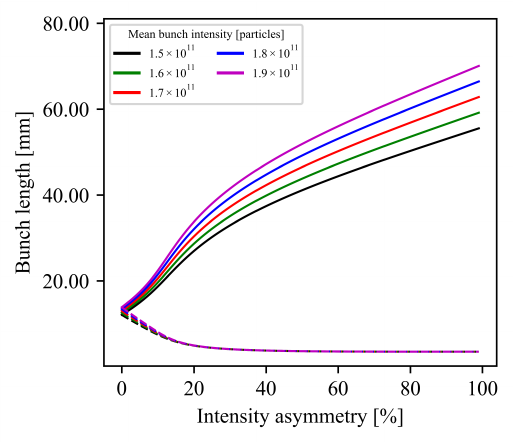}
		}
		\subfloat[Z, weak hourglass, no crab waist]{
			\includegraphics[width=.475\linewidth]{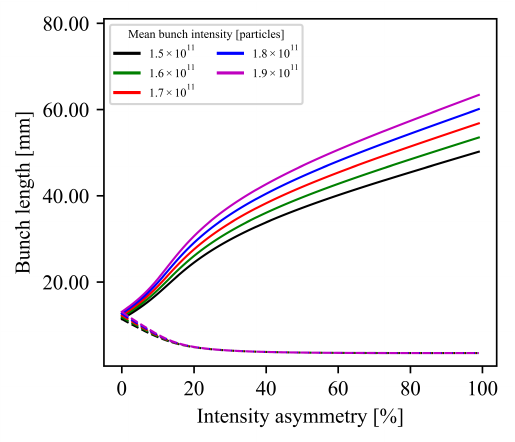}
		}
		\newline
		\subfloat[WW, no hourglass, no crab waist]{
			\includegraphics[width=.475\linewidth]{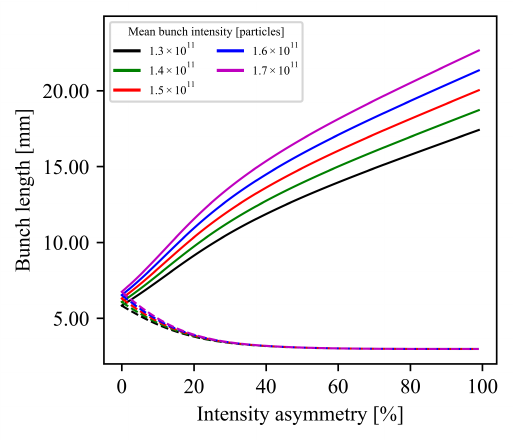}
		}
		\subfloat[WW,  weak hourglass, weak crab waist]{
			\includegraphics[width=.475\linewidth]{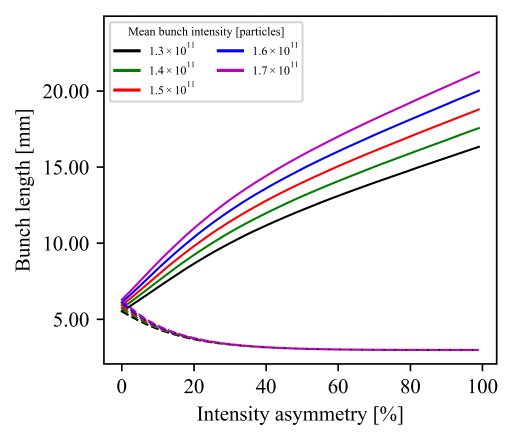}
		}
		\newline
		\subfloat[ZH, no hourglass, no crab waist]{
			\includegraphics[width=.475\linewidth]{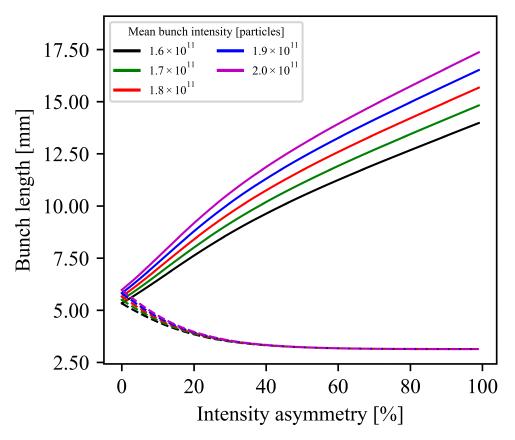}
		}
		\subfloat[ZH, weak hourglass, weak crab waist]{
			\includegraphics[width=.475\linewidth]{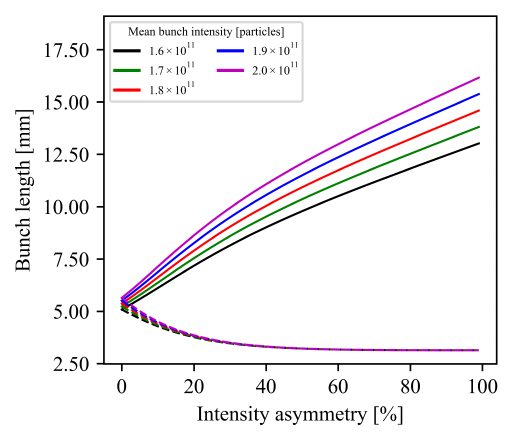}
		}
		\caption{(Caption on next page)}
	\end{figure}
	\begin{figure}
		\ContinuedFloat
		\subfloat[First t\=t, no hourglass, no crab waist ]{
			\includegraphics[width=.475\linewidth]{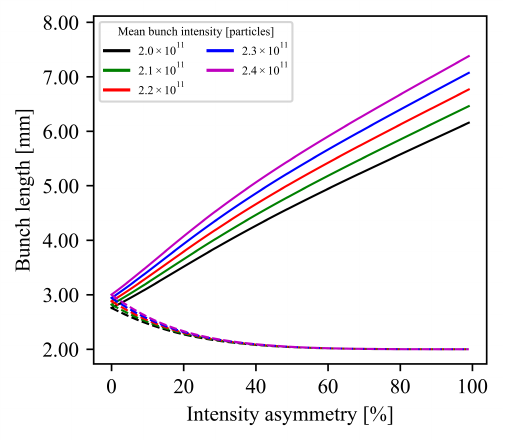}
		}
		\subfloat[First t\=t, weak hourglass, weak crab waist]{
			\includegraphics[width=.475\linewidth]{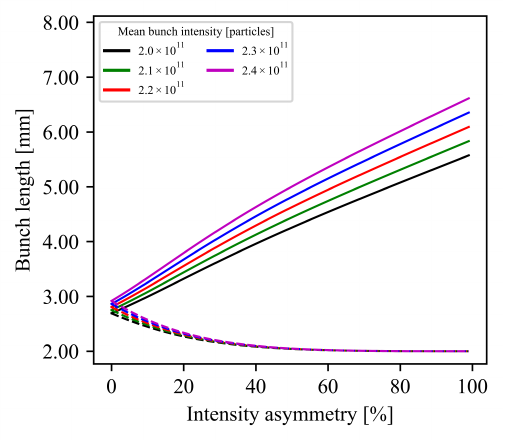}
		}
		\newline
		\subfloat[Second t\=t, no hourglass, no crab waist ]{
			\includegraphics[width=.475\linewidth]{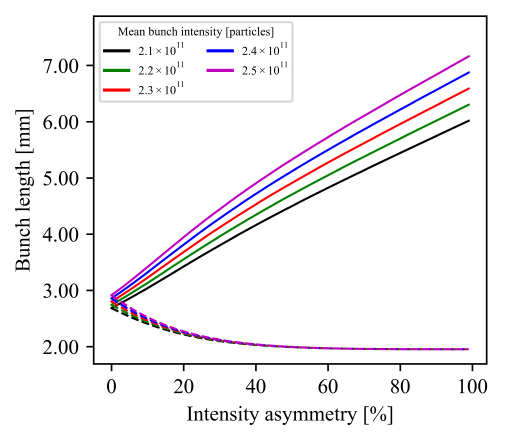}
		}
		\subfloat[Second t\=t, weak hourglass, weak crab waist]{
			\includegraphics[width=.475\linewidth]{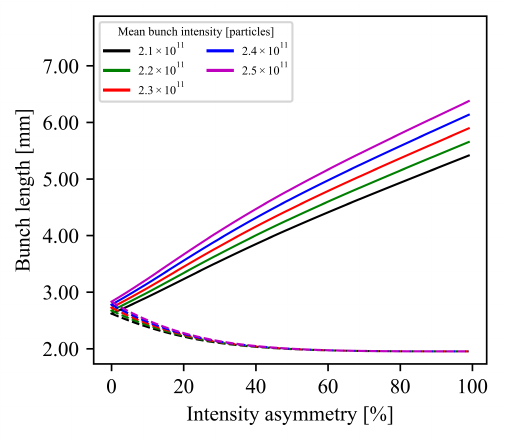}
		}
		\caption{\label{appfig:bunch-length-1d}Equilibrium bunch lengths determined by applying the different approximations of the hourglass and crab waist strength to the 1D model at various mean intensities centred around the intended value in Tab.~ I
		. Other machine parameters in Tab.~I 
		were used for all resonances. The solid line shows the weak bunch length, and strong bunch length is plotted as the broken line. The red line corresponds to the proposed mean intensity.}
	\end{figure}
	
	\begin{figure}[!h]
		\centering
		\subfloat[Z, $\lambda_{++}$]{
			\includegraphics[width=.475\linewidth]{pdf/eigenvalues_Z_plusplus_1D_nohgnocw_19Jan2023.pdf}
		}
		\subfloat[Z, $\lambda_{+-}$]{
			\includegraphics[width=.475\linewidth]{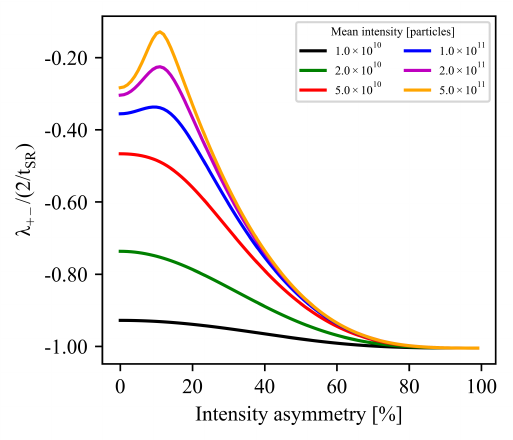}
		}
		\newline
		\subfloat[WW, $\lambda_{++}$]{
			\includegraphics[width=.475\linewidth]{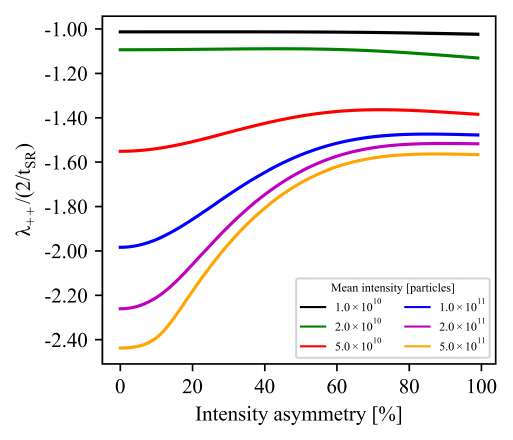}
		}
		\subfloat[WW, $\lambda_{+-}$]{
			\includegraphics[width=.475\linewidth]{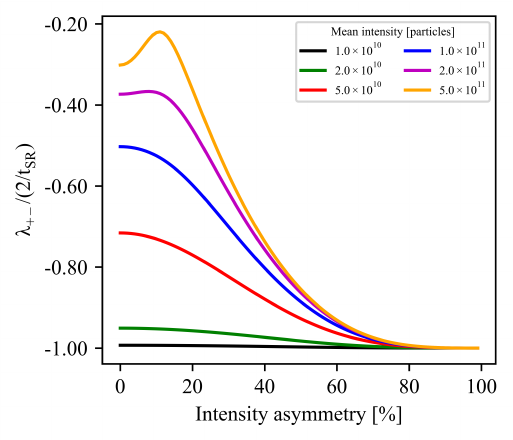}
		}
		\newline
		\subfloat[ZH, $\lambda_{++}$]{
			\includegraphics[width=.475\linewidth]{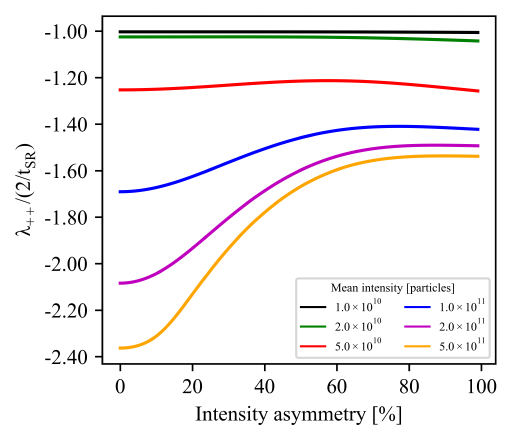}
		}
		\subfloat[ZH, $\lambda_{+-}$]{
			\includegraphics[width=.475\linewidth]{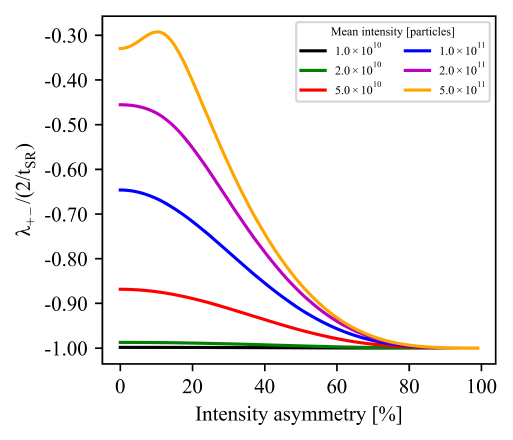}
		}
		\caption{(Caption on next page)}
	\end{figure}
	\begin{figure}
		\ContinuedFloat
		\subfloat[First t\=t, $\lambda_{++}$]{
			\includegraphics[width=.475\linewidth]{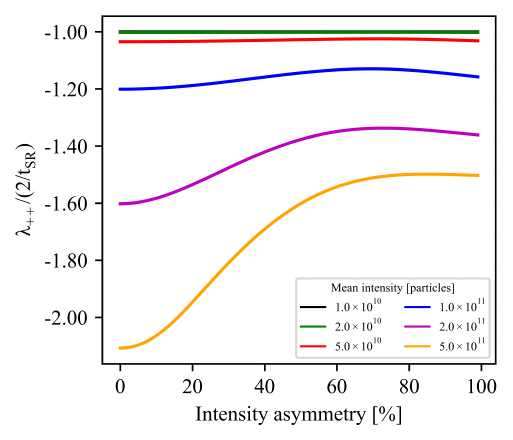}
		}
		\subfloat[First t\=t, $\lambda_{+-}$]{
			\includegraphics[width=.475\linewidth]{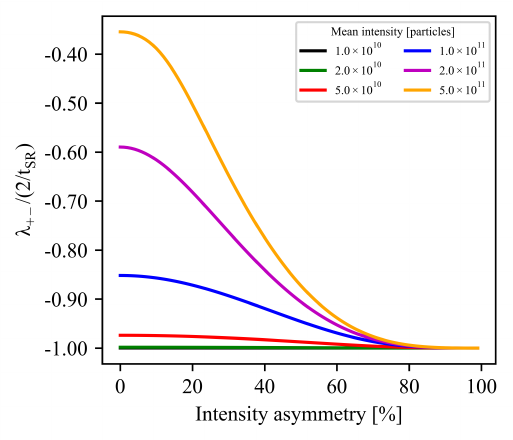}
		}
		\newline
		\subfloat[Second t\=t, $\lambda_{++}$]{
			\includegraphics[width=.475\linewidth]{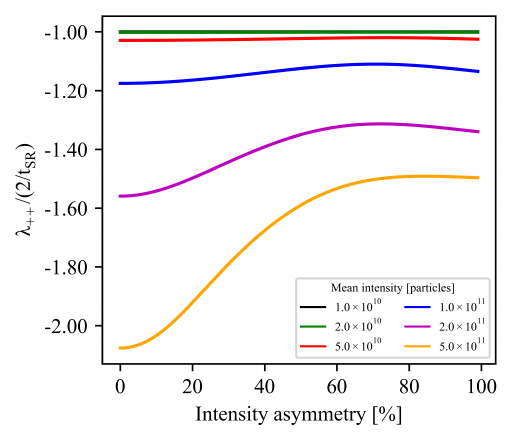}
		}
		\subfloat[Second t\=t, $\lambda_{+-}$]{
			\includegraphics[width=.475\linewidth]{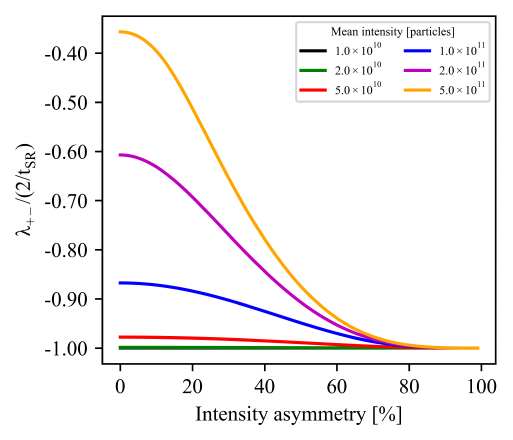}
		}
		\caption{\label{appfig:eig-1D-nohgnocw}Eigenvalues determined when applying linear stability analysis to the equilibrium bunch lengths obtained under the 1D ``no hourglass, no crab waist'' approximation using machine parameters corresponding to various resonances.}
	\end{figure}
	\pagebreak
	\begin{figure}[!h]
		\centering
		\subfloat[Z, no hourglass, no crab waist]{
			\includegraphics[width=.475\linewidth]{pdf/eqm_bunch_lengths_Z_resonance_3D_nohgnocw_08Aug2023}
		}
		\subfloat[Z, weak hourglass, no crab waist]{
			\includegraphics[width=.475\linewidth]{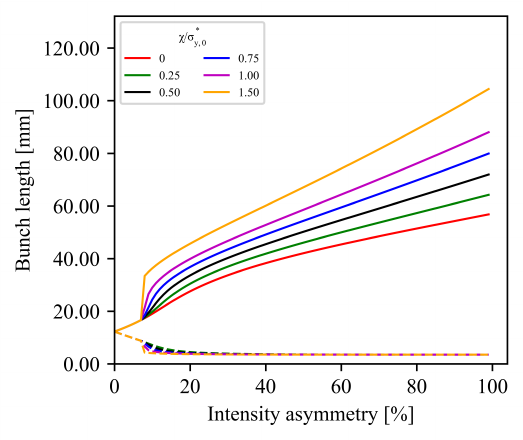}
		}
		\newline
		\subfloat[WW, no hourglass, no crab waist]{
			\includegraphics[width=.475\linewidth]{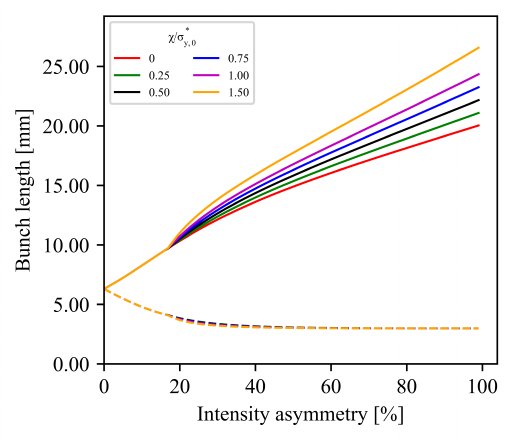}
		}
		\subfloat[WW,  weak hourglass, weak crab waist]{
			\includegraphics[width=.475\linewidth]{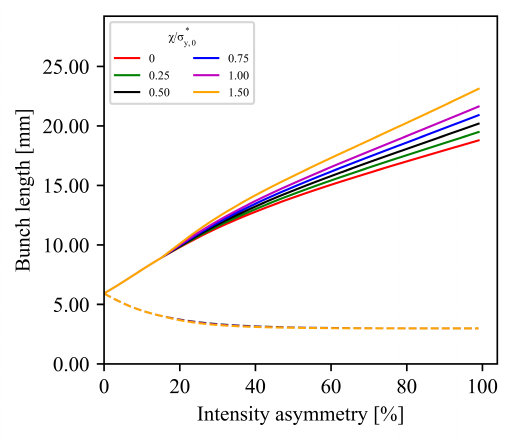}
		}
		\newline
		\subfloat[ZH, no hourglass, no crab waist]{
			\includegraphics[width=.475\linewidth]{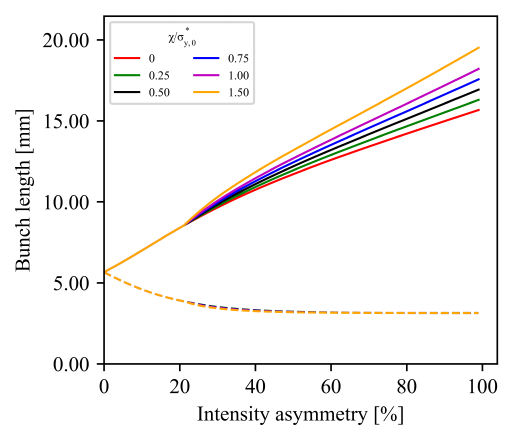}
		}
		\subfloat[ZH, weak hourglass, weak crab waist]{
			\includegraphics[width=.475\linewidth]{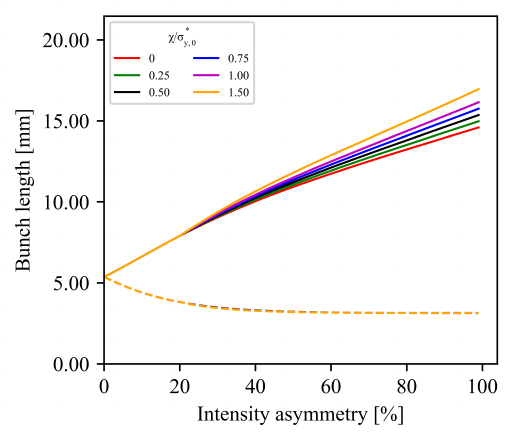}
		}
		\caption{(Caption on next page)}
	\end{figure}
	\begin{figure}
		\ContinuedFloat
		\subfloat[First t\=t, no hourglass, no crab waist ]{
			\includegraphics[width=.475\linewidth]{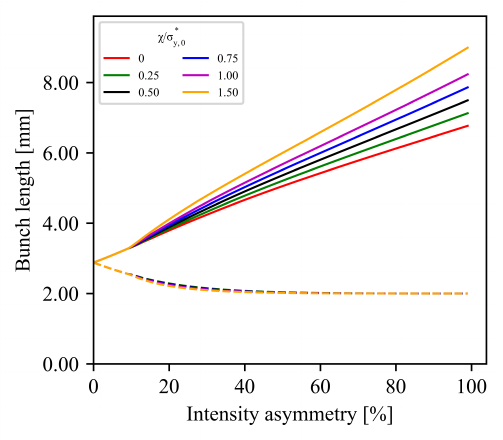}
		}
		\subfloat[First t\=t, weak hourglass, weak crab waist]{
			\includegraphics[width=.475\linewidth]{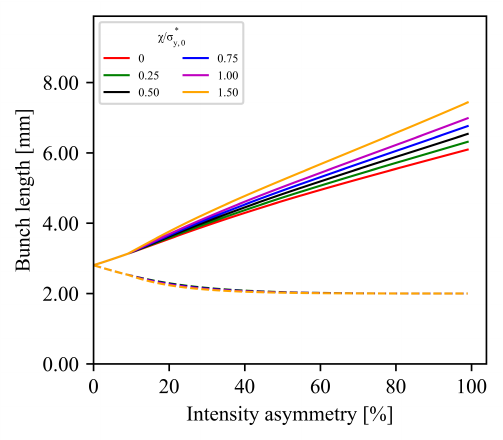}
		}
		\newline
		\subfloat[Second t\=t, no hourglass, no crab waist ]{
			\includegraphics[width=.475\linewidth]{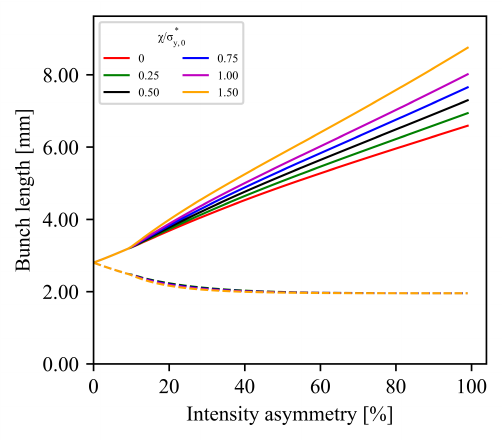}
		}
		\subfloat[Second t\=t, weak hourglass, weak crab waist]{
			\includegraphics[width=.475\linewidth]{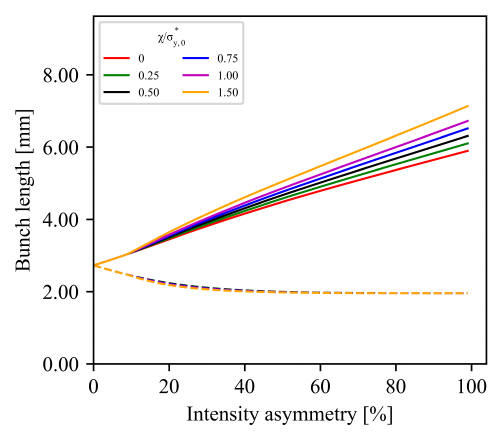}
		}
		\caption{\label{appfig:3d-length-width-nohgnocw}Equilibrium bunch lengths, $\sigma_{z,w/s,\text{eqm}}$, determined by applying the different approximations of the hourglass and crab waist strength to the phenomenological 3D model with threshole beam-beam parameter $\xi_0 = 0.20$ and various values of the transverse blow-up parameter $\chi$. Machine parameters in Table I
		were used for all resonances. The solid line shows the weak bunch length, and strong bunch length is plotted as the broken line. The red line, with $\chi=0$, is the case where there is no $zy-$coupling, \emph{i.e.} the 1D model.}
	\end{figure}
	\pagebreak
	\clearpage 
	\bibliography{bibliography}